\documentclass[aps,prr,reprint,times,floatfix,superscriptaddress]{revtex4-2}

\usepackage{amsmath}
\usepackage{amssymb}
\usepackage{bm}
\usepackage{graphicx,subfigure}
\usepackage{epsf,epsfig}
\usepackage{hyperref}
\usepackage{xcolor}
\usepackage{accents}
\usepackage{soul}

\DeclareMathOperator{\sech}{sech}

\newcommand{\ai}[1]{\textcolor{black}{#1}}

\newcommand{\sh}[1]{\textcolor{black}{#1}}

\begin{document}

\title{Exotic rheology of materials with active rearrangements}

\author{Aondoyima Ioratim-Uba}
\affiliation{School of Mathematics, University of Bristol - Bristol BS8 1UG, UK}
\affiliation{Max Planck Institute for Dynamics and Self-Organization (MPI-DS), D-37077 G{\"o}ttingen, Germany}
\affiliation{Living Systems Institute, University of Exeter - Exeter EX4 4PY, UK}
\author{Tanniemola B. Liverpool}
\affiliation{School of Mathematics, University of Bristol - Bristol BS8 1UG, UK}
\affiliation{Isaac Newton Institute for
Mathematical Sciences, 20 Clarkson Rd, Cambridge CB3 0EH, UK}
\author{Silke Henkes}
\affiliation{School of Mathematics, University of Bristol - Bristol BS8 1UG, UK}
\affiliation{Lorentz Institute for Theoretical Physics, LION, Leiden University - Leiden 2333 CA, The Netherlands}
\affiliation{Isaac Newton Institute for
Mathematical Sciences, 20 Clarkson Rd, Cambridge CB3 0EH, UK}

\date{\today}

\begin{abstract}
The flow of biological tissues during development is controlled through the active stresses generated by cells interacting with their mechanical environment in the tissue. Many developmental processes are driven by convergence-extension flows where the tissue has an emergent negative shear modulus and viscosity. This exotic rheology is generated through active T1 transitions where rearrangements are opposite the applied stress direction. Here, we introduce a mean-field elasto-plastic model which shows convergence-extension, based on the Hebraud-Lequeux model, that includes both passive and active elastic elements with opposite stress responses. 
We find that the introduction of active elements profoundly changes the rheology. Beyond a threshold fraction of active elements, it gives rise to non-monotonic flow curves and negative stresses at positive strain rates. Controlled by the active fraction and the stress diffusion rate, we find both yield stress materials and fluids, with either a positive or negative yield stress or viscosity. \sh{Furthermore, allowing for residual active stresses generates rheological curves with an asymmetry in strain rate dependence, resembling hysteresis.} These features are characteristic of metamaterials, and highlight how biology uses disordered, active materials with exotic rheology.  
\end{abstract}

\maketitle

\section{Introduction}
Tissues are active viscoelastic materials \cite{petridou2019tissue, guillot2013mechanics, collinet2015local} with complex rheological properties ~\cite{Hatwalne04,Liverpool2006b,Giomi2010,Loisy2018,Loisy2019} that depend on cellular composition \cite{park2016cell,atia2018geometric}, the mechanical environment \cite{sonam2023mechanical,storm2005nonlinear,engstrom2019compression}, and internal active stresses \cite{rozbicki2015myosin,mongera2018fluid,jain2024cell}. Theoretical models have captured many of these rheological properties including yield stress behaviour \cite{hopkins2022local, popovic2021inferring, nguyen2025origin,grossman2022instabilities}, and transitions between fluid and solid behaviour controlled by single cell properties \cite{bi2015density,bi2016motility} and external shear \cite{huang2022shear,fielding2023constitutive,anand2026non}.

\begin{figure*}
	\centering
	\includegraphics[width=1.0\textwidth]{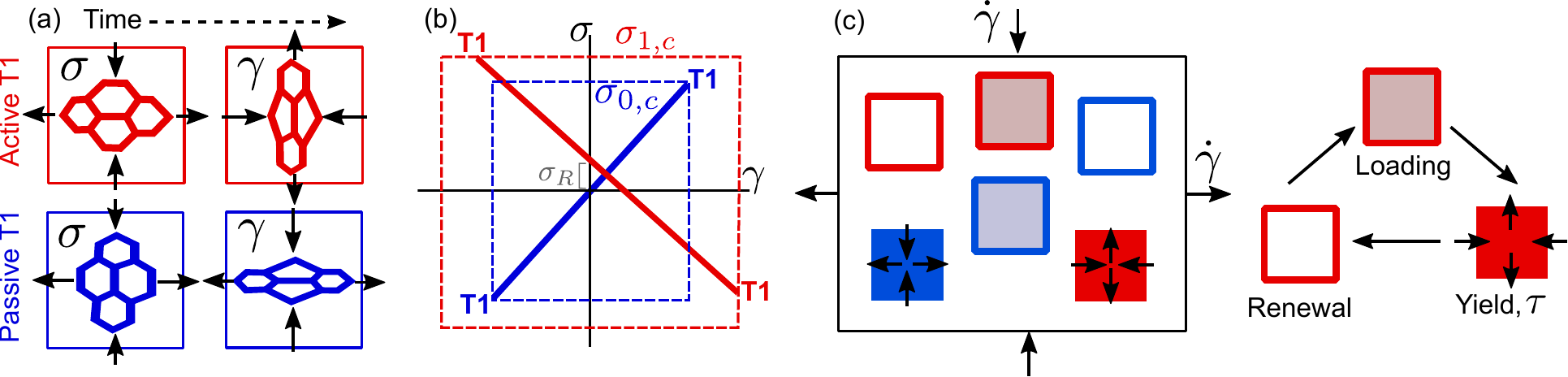}
	\caption{(a) Illustration of an active T1 (red) and a passive T1 (blue) event within a single element. Stress has the opposite sign to strain in active T1s. (b) Elasto-plastic response. Red represents an active element, and blue represents a passive block. Stress $\sigma$ depends linearly on strain up to the yield stress $\sigma_{k,c}$, where a T1 is triggered. \ai{We also consider an active population with a residual stress $\sigma_R$, shown as an offset along the stress axis.} (c) Possible states of a block in a material under strain rate $\dot{\gamma}$: elastic loading (partially filled square), yielding with time scale $\tau$ (filled square with arrows), renewal (empty square).}
	\label{fig:1}
\end{figure*}

Convergence-extension (CE) is a conserved morphogenic process where a region of epithelial tissue elongates in one direction and contracts along a different direction.
This process is controlled by active cell neighbour exchanges known as active T1 transitions \cite{rozbicki2015myosin, krajnc2018fluidization} (FIG. \ref{fig:1}. (a)), named in analogy to the well-known topological transitions that relax stress in passive foams in response to external driving \cite{graner2008discrete}. In tissues, however, active stresses are generated internally by the actomyosin cortex, without external driving, thus making T1s active. 

Convergence-extension rheology is controlled by the rate and spatial distribution of active T1s~\cite{huebner2018coming,sutherland2020convergent}. The mechanism is that junctions that align with the (chemical or mechanical) polarisation axis of the tissue preferentially contract, developing local stresses that oppose external boundary forces \cite{collinet2015local}. They then realign \emph{opposite} the direction of stress  \cite{rauzi2008nature,rauzi2015embryo}, a clearly active process that has no equivalent in passive rheology. Thus active T1s are central to our understanding of tissue flow and large scale deformation \cite{rozbicki2015myosin,tetley2018same,jain2024cell,aboutaleb2023random_yielding}.

This exotic active rheology has features reminiscent of the physics of metamaterials, with their coupling between local elements and an emergent auxetic or otherwise unusual response, in both engineered \cite{dudek2025shape} and disordered materials \cite{zaiser2023disordered}. This includes for example a negative response to applied stress \cite{ducarme2025exotic}, and emergent active locomotion \cite{veenstra2025adaptive} through exploiting the properties of odd active materials \cite{fruchart2023odd}. Yet, these materials are only beginning to be constructed at the microscale \cite{melio2025colloidal}.  Furthermore, the biological version rearranges and rebuilds its mechanical structure, a feat yet to be achieved by engineered systems.

Whilst the full picture remains unclear, a number of models for convergence-extension have begun to emerge, which couple  junction dynamics to a stress field~\cite{popovic2017active,duclut2022active} and incorporate mechanical feedback, in systems with rotational symmetry externally~\cite{rauzi2008nature,staddon2019mechanosensitive} or spontaneously broken~\cite{sknepnek2021generating,ioratim2023mechano}. By incorporating mechano-chemical feedback in a vertex model one can generate a detailed model for the dynamics of feedback in active T1s and achieve spontaneous tissue convergence extension \cite{sknepnek2021generating}. Stress redistribution couples T1s, and tissue-scale tension chains strongly influence the process. Using the same feedback mechanism in a continuum model it is possible to achieve robust CE flows through a pitchfork bifurcation~\cite{ioratim2023mechano}. Most recently, data driven modeling has achieved convergence-extension through feedback in a stress-based formulation \cite{brauns2024geometric,claussen2024geometric,claussen2024mean}. All these approaches show that junction orientations and network topology play a significant role.
Long-range coordination via stress redistribution of T1s is therefore crucial to understand the systematic tissue elongation we see in CE. 

\ai{These considerations naturally motivate the use of elasto-plastic models for disordered materials \cite{nicolas2018deformation}. These models bridge between the microscopic and the macroscopic by defining the stress for discrete mesoscopic blocks in the material that can support exactly one yield event at a time.} A single active T1 is then a yield event that can trigger other active T1s by altering the stress felt by other parts of the tissue \cite{jain2024cell, duclut2022active}. These models must also account for the transmission of mechanical fluctuations since a single block feels the effect of several other blocks at once. The simplest way for this is through mechanical noise and then to use an effective diffusion coefficient in a mean-field approach \cite{puosi2015probing}. 
The aspects of tissue rheology that have been captured so far by mean-field approaches include a finite yield stress \cite{fielding2023constitutive,nguyen2025origin}, and tissue fluidization \cite{matoz2017nonlinear} and stress equilibria through cell division and apoptosis \cite{tahaei2025cell,tahaei2026celldivisionssuppressdynamical}. However, the mechanics of feedback active T1 transitions that yield opposite to polarisation has not been explored.

\sh{Here, we} investigate the flow properties of a {\emph{two component}} disordered material with passive and active elements using a \sh{mean-field} elasto-plastic model. The key new feature of our model that leads to novel anomalous rheology~\cite{matoz2017nonlinear} is the presence of a new mesoscale active component with opposite stress-strain relations.  \sh{To allow for tractability, we restrict ourselves to a one-dimensional shear stress, and a fixed fraction of active and passive components.} We obtain \ai{closed-form} expressions for the macroscopic stress and the yield stress as a function of applied strain rate and find non-monotonic flow curves when there are enough, but not too many, active elements in the material. \sh{Such non-monotonic flow curves are typical of complex materials such colloidal suspensions or foams. However, in our active case we find that the mean stress is of opposite sign to the strain at low strain rate, and recovers the same sign at high strain rate.} We also see the emergence of a negative yield stress when the stress diffusion rate is low and there is a sufficient fraction of active elements. At higher stress diffusion rates, the material is fluid. \ai{ We explain this behaviour in terms of competition between the fraction of active elements and the difference in the yield stresses of the active and passive components.}
We summarize the response of the material in a phase diagram characterized by the sign of the yield stress and the derivative of macroscopic stress with respect to strain rate, \sh{i.e. of the effective viscosity in the limit of small and large strain rates}. \sh{We furthermore investigate the effect of a non-zero residual active stress after yielding, which breaks symmetry under strain rate reversal. The resultant flow curves are also asymmetric, with positive or negative stress in the material at zero strain rate, in a hysteresis-like effect.}
\sh{In summary, allowing for active elements that yield opposite the applied stress gives rise to a variety of exotic flow curves, controlled by the active yield and residual stresses, the active element fraction and the overall stress diffusion rate.}

\section{Active-passive two component model} 
We develop a two component version of the H\'ebraud-Lequeux (HL) model, which describes disordered passive materials \cite{hebraud1998mode} subject to external strain rate $\dot{\gamma}$. Our core modification is that we include a mixture of active and passive blocks (FIG. \ref{fig:1}. (a)), which requires two probability distributions for the evolution of the stress $\sigma$, $P_1(\sigma)$ for active and $P_0(\sigma)$ for passive blocks,
\begin{equation} \label{eq:probevol}
\begin{split}
\partial_{t}P_k(\sigma,t) = & -G_k\dot{\gamma}\partial_{\sigma}P_k(\sigma,t) + D(t)\partial_{\sigma}^2P_k(\sigma,t) \\ &  - \frac{\theta(|\sigma| - \sigma_{k,c})}{\tau}P_k(\sigma,t) + \delta(\sigma)\phi_k\Gamma(t).
\end{split}
\end{equation}
Here $k = 0,1$ labels passive and active elements, and $0 < \phi < 1$ is the active fraction of the system, i.e. $\phi_1 = \phi$ and $\phi_0=1-\phi$. The first terms on the RHS of \eqref{eq:probevol} encodes the elastic response represents advection in stress space of the stress distribution by $G_i\dot{\gamma}$.
Passive blocks have stress increasing with strain $\sigma = G_0\gamma$, with \sh{elastic modulus} $G_0>0$ up to a local yield stress $\sigma_{0,c}$, modeling passive, foam-like T1 transitions.
In contrast, active elements have stress opposite to strain, $\sigma = G_1\gamma$ with \sh{effective active modulus} $G_1<0$, up to a local yield stress $\sigma_{1,c}$, modeling feedback active T1 transitions. Beyond its yield stress, a block yields through an active or passive T1 transition (FIG. \ref{fig:1}. (b)).

\begin{figure}[t!]
	\includegraphics[width=\columnwidth]{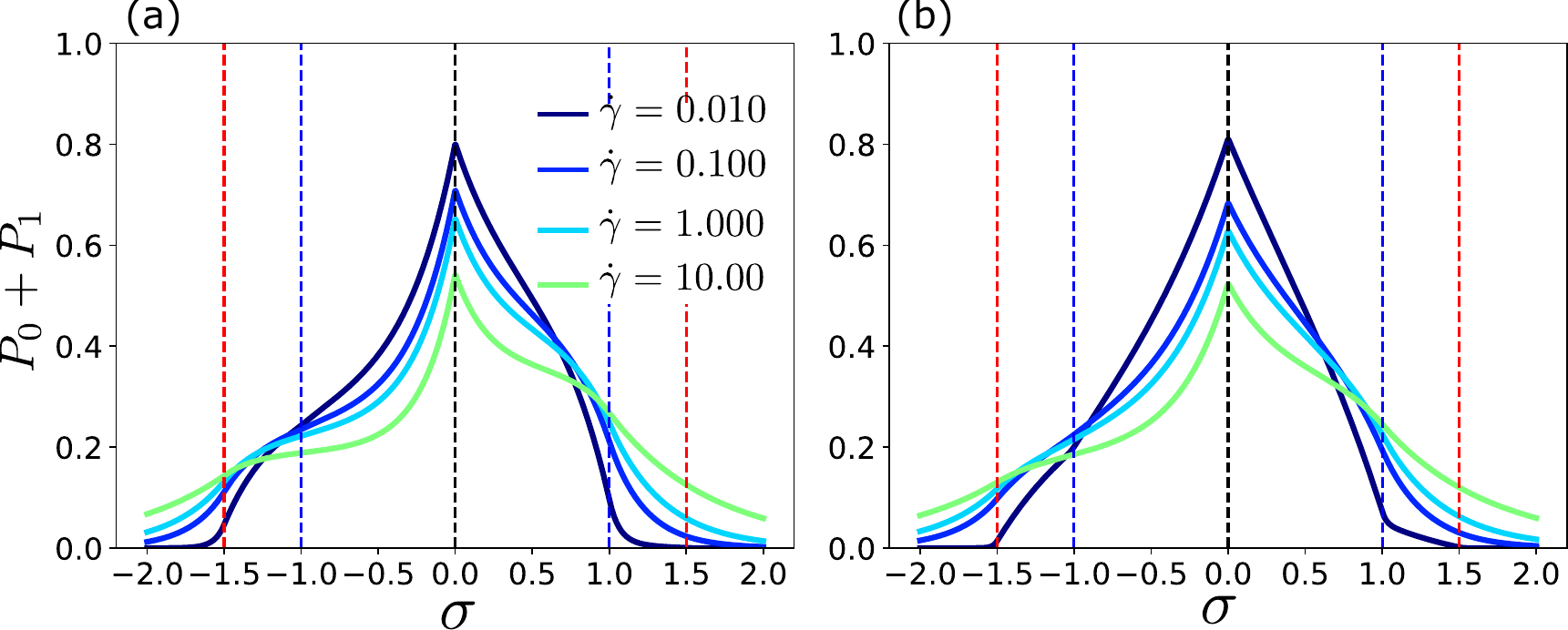}
	\caption{\ai{Probability distributions for $\phi = 0.35$ at (a) $\alpha = 0.4$, and (b) $\alpha = 0.6$. The dashed red lines show the yield stress for active elements $\pm\sigma_{1,c}$, and the dashed blue lines show the yield stress for passive elements $\pm\sigma_{0,c}$. The dashed black line shows $\sigma = 0$.}}
	\label{fig:1-5}
\end{figure} 

\begin{figure*}
	\centering
	\includegraphics[width=1.0\textwidth]{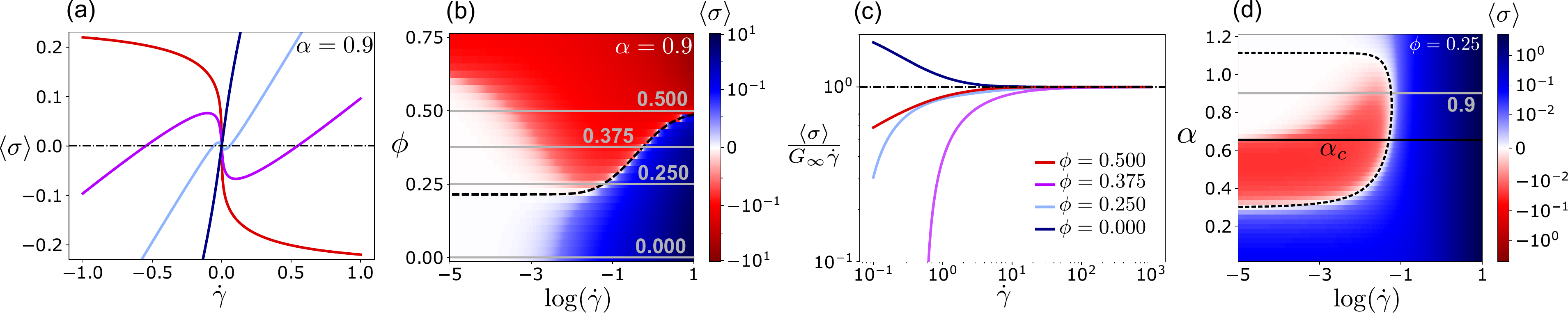}
	\caption{(a) Macroscopic stress $\langle \sigma \rangle$ against strain rate $\dot{\gamma}$ at fixed $\alpha = 0.9$ for $\phi = 0, 0.25, 0.375, 0.5$. (b) Dependence of $\langle \sigma \rangle(\dot{\gamma})$ on active fraction $\phi$ at fixed $\alpha = 0.9$. The dashed black line is $\langle \sigma \rangle = 0$. The \ai{grey} lines correspond to the curves in (a). (c) The large strain rate behaviour is $\langle \sigma \rangle = G_{\infty}\dot{\gamma}$ or \ai{$\langle \sigma \rangle = (\sigma_{0,c} - \sigma_{1,c})/2$ for $\phi = 0.5$} (equations \eqref{eq:large_strain_rate} and \ai{\eqref{eq:large_stress_equal_mix}}). (d) Dependence of $\langle \sigma \rangle (\dot{\gamma})$ on $\alpha$ at fixed $\phi = 0.25$. The solid black line shows $\alpha_c$ and the green line shows $\alpha = 0.9$.}
	\label{fig:2}
\end{figure*}

The second term encodes the mechanical noise from stress redistribution in an effective diffusivity $D(t)$.  The third term on the RHS of \eqref{eq:probevol} determines what happens when the stress magnitude in a block exceeds $\sigma_{k,c}$, indicated by the dashed box in FIG. \ref{fig:1}. (b). It corresponds to the removal of over-stressed blocks over a time scale $\tau$, where $\theta(x) = 1$ for $x > 0$ and is zero otherwise. The last term corresponds to the addition of new elements at zero stress at the same rate that yield events are occurring - this rate is $\phi_k\Gamma(t)$.
Self-consistency requires it to be proportional to the overall rate of yielding events in the material, the plastic activity $\Gamma(t)$, i.e. $D(t) = \alpha \Gamma(t)$ with
\begin{equation}
\Gamma(t)\tau = \int\limits_{|\sigma| > \sigma_{0,c}} P_0(\sigma,t) \,d\sigma + \int\limits_{|\sigma| > \sigma_{1,c}} P_1(\sigma,t) \,d\sigma,
\end{equation}
where $\tau$ is the time scale of a single plastic event. The temperature-like parameter $\alpha$ tunes the mechanical propagation of yielding events and controls the transition from a yield stress material to a fluid.
FIG. \ref{fig:1}. (c) illustrates the ensemble of states that passive and active elements are in while the material is subject to a pure strain rate $\dot{\gamma}$. 

We seek the stationary solution $\partial_tP_k = 0$, which gives rise to a second order differential equation. Each distribution can be solved for independently, assuming a fixed but unknown $D$ (see SI). The active and passive blocks are coupled by the fact that the effective diffusion is constant throughout the material. Once we have solved for $P_k(\sigma)$ in terms of $D$, we require that $\int_{-\infty}^{\infty} (P_0 + P_1) \,d\sigma = 1$, and combine this with the self-consistency requirement to obtain an implicit equation for $D$. First, the normalisation condition gives an equation of the form $\Gamma\tau = D\tau/f_{\alpha}$, where $f_{\alpha}\left(\sqrt{D\tau},G_0\dot{\gamma}\tau,G_1\dot{\gamma}\tau\right)$ is a function determined by the form of $P_0(\sigma)$ and $P_1(\sigma)$. Then, invoking equation $D(t) = \alpha \Gamma(t)$ gives us the implicit equation for $D$
\begin{equation} \label{eq:D_eqn}
f_{\alpha}\left(\sqrt{D\tau},G_0\dot{\gamma}\tau,G_1\dot{\gamma}\tau\right) - \alpha = 0.
\end{equation}
The physical system we are modelling is a material subject to macroscopic strain rate $\dot{\gamma}$, in which the corresponding steady state macroscopic stress is given by
\begin{equation} 
\langle \sigma \rangle = \int_{-\infty}^{\infty} \sigma P_0(\sigma) \,d\sigma + \int_{-\infty}^{\infty} \sigma P_1(\sigma) \,d\sigma.
\end{equation}
\ai{The full probability distribution $P_0 + P_1$ for $\phi = 0.35$ is shown in FIG. \ref{fig:1-5}}.

\section{Results} 
We examine the resulting flow curves, macroscopic stress as a function of strain rate $\langle \sigma \rangle (\dot{\gamma})$, that characterise the material (FIG. \ref{fig:2}. (a)). \sh{We examine the macroscopic response at low as well as large strain rate through both the sign of the macroscopic stress, and the sign of the gradient of the stress $\partial_{\dot{\gamma}}\langle \sigma \rangle$ relative to the strain rate $\dot{\gamma}$, i.e. the incremental viscosity.} 
For simplicity, we focus on a system where $|G_0| = |G_1|$, and we choose $\sigma_{1,c}/\sigma_{0,c} > 1$, that is an active local yield stress that is larger than the passive one. \sh{The opposite choice $\sigma_{1,c}/\sigma_{0,c} < 1$ with a lower active yield stress leads to monotonic flow curves without interesting properties (see FIG.~\ref{SI_fig:3.5} (b)).} Note that exactly equal and opposite material properties lead to a trivial solution where $\langle \sigma \rangle = (1-2\phi) \langle \sigma \rangle_{HL} $, the standard H\'ebraud-Lequeux stress curves. Throughout this section, we fix the parameters $G_0 = 1$, $G_1 = -1$, $\tau = 1$, $\sigma_{0,c} = 1$, $\sigma_{1,c} = 1.5$. The \ai{closed-form} expression for $\langle \sigma \rangle$ can be found in the SI (eq. S9). 


At large strain rate, we find that the sign of the macroscopic stress \ai{and its gradient} is governed by $\phi$ for fixed $G_0$ and $G_1$. Taking the limit $\dot{\gamma} \to \infty$, we find
\ai{
\begin{equation} \label{eq:large_strain_rate}
\lim_{\dot{\gamma} \to \infty}\frac{\langle \sigma \rangle}{\dot{\gamma}} =  \big[(1 - \phi)G_0 - \phi\left|G_1\right|\big]\tau \equiv G_{\infty},
\end{equation}
}
\ai{where $G_{\infty}$ is the effective viscosity at large strain rate. The special case $\phi = 1/2$ and $|G_1| = G_0$ must be done separately, and we find:
\begin{equation} \label{eq:large_stress_equal_mix}
\lim_{\dot{\gamma} \to \infty}\langle \sigma \rangle = \frac{1}{2}\big(\sigma_{0,c} - \sigma_{1,c}\big).
\end{equation}
} FIG. \ref{fig:2}. (c) shows that $\langle \sigma \rangle/(G_{\infty}\dot{\gamma}) \to 1$ (or for $\phi = 0.5$, \ai{$2\langle \sigma \rangle/(\sigma_{0,c} - \sigma_{1,c}) \to 1$}) as the strain rate becomes large (see SI figure S2 (a) for a larger range of $\phi$), and demonstrates that equation \eqref{eq:large_strain_rate} determines the sign of the stress for $\dot{\gamma} \gtrsim 1$~\footnote{At $\phi = 0.5$, the limit $\dot{\gamma} \to 0$ is singular, and one must be careful taking this limit (see SI FIG. S2. (b)).}. \sh{Physically, to ensure the energy input into the system remains finite, one requires that response at large strain rate be passive, that is the material needs to yield in the direction of applied stress so that $G_{\infty}>0$.}

\ai{Our main result is that coupling active and passive yielding events within a mean-field elasto-plastic model generates a rich family of flow curves, including responses characteristic of convergence-extension in which the material yields opposite to the applied stress at low strain rates but recovers passive behaviour at large strain rates.} \ai{That is, there are local minima and maxima in $\langle \sigma \rangle(\dot{\gamma})$, resulting in non-monotonic flow curves} (FIG.~\ref{fig:2}(a)). \ai{In this regime, the macroscopic response changes with strain rate: at low strain rates the stress can oppose the imposed deformation, whereas at high strain rates it becomes aligned with it.} This behaviour requires a balance between active and passive yielding elements - \ai{with mostly passive elements the macroscopic stress remains aligned with the imposed strain rate across all $\dot{\gamma}$, whereas with mostly active elements the response can remain opposed across all strain rates. In the intermediate regime, the system crosses between these two behaviours, producing non-monotonic flow curves.} In FIG. \ref{fig:2}. (b), we show a colour plot of stress as a function of positive strain rate and of $\phi$, with positive stresses in blue and negative ones in red. The appropriate range is above the dashed $\langle \sigma \rangle=0$ line until $\phi=0.5$, where the system turns fully active. When the system is purely passive ($\phi = 0$), we recover the flow curves of the original HL model with passive response for all $\dot{\gamma}$.

As in the original HL model, our model has a transition from a yield stress material to a fluid that is controlled by the parameter $\alpha$. FIG.  \ref{fig:2}. (d) shows the stress curves $\langle \sigma \rangle(\dot \gamma)$ depend on $\alpha$, starting from a material with a positive yield stress $\langle \sigma \rangle(0)$, then switching to a negative yield stress (but passive flow at larger $\dot \gamma$), and finally no yield stress above $\alpha_c$. The critical value $\alpha_c$ is obtained by writing down $\partial_t P_k = 0$ at zero strain rate and looking for the value of $\alpha$ below which there are no real solutions for $D$:
\begin{equation} \label{eq:alpha_crit_main}
\alpha_c = \left[(1 - \phi){\sigma_{0,c}}^2 + \phi{\sigma_{1,c}}^2\right]\big/2.
\end{equation}
We obtain a \ai{closed-form} prediction for the yield stress and the low strain rate behaviour of fluid state by taking the limit $\dot{\gamma} \to 0$. Below $\alpha_c$, in the yield stress state, we have a Herschel-Bulkley law with an exponent of $\frac{1}{2}$, 

\begin{equation}
\langle\sigma\rangle \simeq \sigma_Y(\phi, \sigma_{0,c}, \sigma_{1,c}) + \tilde{A}(\phi, \sigma_{0,c}, \sigma_{1,c})(\tau\dot{\gamma})^{\frac{1}{2}},
\end{equation}
and above $\alpha_c$, in the fluid state, we have that stress is linear in strain rate:
\begin{equation}
\langle\sigma\rangle \simeq (1 - \phi)A_0(\sigma_{0,c})G_0\tau\dot{\gamma} - \phi A_1(\sigma_{1,c})|G_1|\tau\dot{\gamma},
\end{equation}
where the expressions for the yield stress $\sigma_Y$ and the power law coefficients $\tilde{A}, A_0$, and $A_1$ are given in the SI in equations S19 - S24. These results have the same form as the equivalent results for the original HL model, as in \cite{agoritsas2015relevance}, and reduce to the results for the original model for $\phi = 0$. The key difference is that the yield stress and power law coefficients can become negative.

\begin{figure*}[t!]
	\centering
	\includegraphics[width=1.0\textwidth]{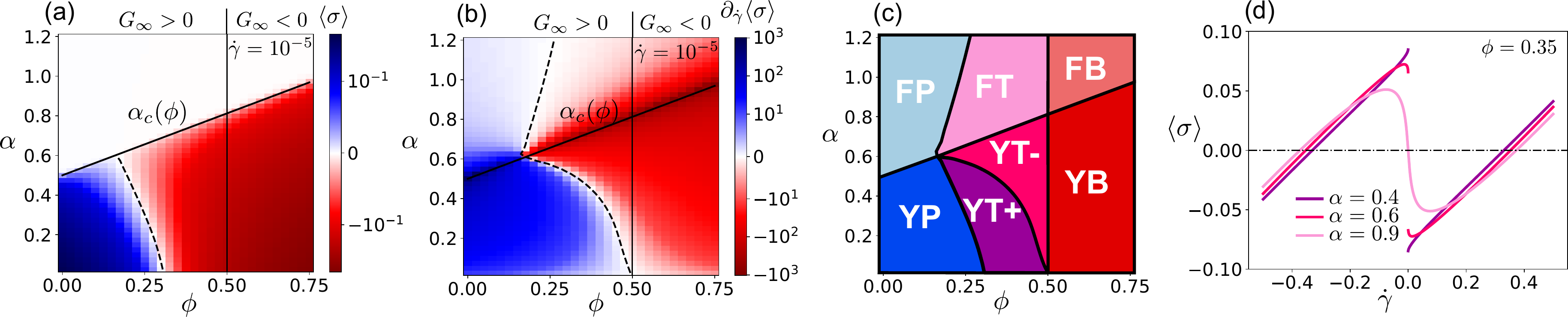}
	\caption{(a) Dependence of the macroscopic stress $\langle \sigma \rangle$ at low strain rate $\dot{\gamma} = 10^{-5}$ on the active fraction $\phi$ and $\alpha$ (obtained numerically). The vertical line is $\phi = 0.5$, the diagonal line is $\alpha_c(\phi)$, and the dashed line is $\eta_0 = 0$. (b) Dependence of the gradient of the macroscopic stress at low strain rate $\dot{\gamma} = 10^{-5}$ on $\phi$ and $\alpha$. The dashed line is $\eta_0 = 0$. (c) Phase diagram in the $(\phi,\alpha)$ plane separating fluid vs solid behavior and passive vs active response. \ai{Each phase is labelled by two letters, with the first letter denoting fluid (F) or yield stress (Y), and the second letter denoting passive(P), backwards (B), or non-monotonic response/sign change of the stress relative to strain rate (T).} (d) Macroscopic stress $\langle \sigma \rangle$ against strain rate $\dot{\gamma}$ at fixed $\phi = 0.35$ representing each turnover state YT+ ($\alpha = 0.4$), YT- ($\alpha = 0.6$), and FT ($\alpha = 0.9$).}
	\label{fig:3}
\end{figure*}

Together with numerics, these expressions allowed us to develop two criteria to label the flow curves: The first is the yield stress $\sigma_Y(\phi, \sigma_{0,c}, \sigma_{1,c})$, shown in FIG. \ref{fig:3}. (a), with a change from positive to negative values as a function of $\phi$ and with $\alpha_c$ clearly visible. The second is the initial effective viscosity $\eta_0 = \lim_{\dot{\gamma}\rightarrow 0} \partial_{\dot \gamma} \langle \sigma \rangle$, shown in FIG. \ref{fig:3}. (b). It also switches from positive to negative, but in different locations. 

Altogether, we identified 7 distinct phases in the $(\phi,\alpha)$ plane by characterising the large strain rate and small strain rate behaviour of the macroscopic stress. These phases are summarised in a phase diagram FIG \ref{fig:3}. (c). 

At low $\phi$, when the material \ai{consists of mostly passive elements}, we have a \ai{typical monotonic passive-like response} for all values of strain rate (P). We define this region to be where $\sigma_{Y} \geq 0$, $\eta_0>0$ and finally $G_{\infty} > 0$. Flow curves in this region do not differ qualitatively from the fully passive case (SI FIG. S3.). We have fluid and yield stress behaviour separated by the line $\alpha_c(\phi)$, with the yield stress phase YP below $\alpha_c$, and the fluid phase FP above.

At larger values of $\phi$ but below $\phi = 0.5$ we have the non-monotonic (T) flow curves that qualitatively resemble convergence-extension \cite{sknepnek2021generating,ioratim2023mechano}. Here, the material \ai{either has a negative yield stress $\sigma_{Y} < 0$ or a negative initial effective viscosity $\eta_0 < 0$, coupled with a positive effective shear modulus $G_{\infty}$ at large strain rate. The non-monotonic flow region is divided into three phases by the line $\alpha_c(\phi)$, and the curve $\eta_0 = 0$ (FIG. \ref{fig:3}. (b)).} Below $\alpha_c$, there are two types of negative yield stress material denoted by YT+ and YT- with $\eta_0>0$ and $\eta_0<0$, respectively, indicating a positive or negative gradient in the flow curve at low strain rate. Above $\alpha_c$, we have a fluid denoted by FT. For this phase, $\eta_0<0$ and the stress is negative at small strain rate but then turns positive. We show representative flow curves in the YT+, YT- and FT regions in FIG \ref{fig:3}. (d).

For a mostly active system $\phi > 0.5$, we have $G_{\infty} < 0$, and negative stress and stress gradient, or backwards flow curves (B). This region has an active response for all values of strain rate, and is separated into in a negative yield stress material YB below $\alpha_c$, and an active fluid FB above $\alpha_c$.

The phenomenology of the turnover phases can be explained in terms of competition between $\phi$ and the difference in yield stresses $\sigma_{1,c} - \sigma_{0,c}$ of the active and passive blocks. For these phases the system is mostly passive, but (as for all phases identified in this publication) the active elements have a higher yield stress. At low strain rates, the flow is backwards because the higher yield stress of the active elements matters more than the fact that the system is more passive: passive blocks yield more readily and their elastic response does not contribute as much to the overall material response compared to the active elements that can load elastically for longer before yielding. At higher strain rates, however, this difference in yield stress becomes negligible and, as we can see from equation \eqref{eq:large_strain_rate}, the behaviour of the system is controlled by $\phi$. 


\section{Effect of residual active stresses}
\sh{Active stresses arising from the cytoskeleton develop in the absence of applied external strain as well. Therefore, we expand our model to include the} \ai{ effect of allowing the active elements to relax to a non-zero residual stress $\sigma_R$ after yielding. \sh{This corresponds to a microscopic active stress-strain relationship $\sigma = \sigma_R + G_1 \gamma$, which should be seen as the two lowest order terms in a series expansion in orders of $\gamma$.} The simplest away to achieve this is to shift the Dirac delta in the equation for the probability distribution governing the active population:
\begin{equation}
\begin{split}
& \partial_{t}P_1(\sigma,t) = -G_1\dot{\gamma}\partial_{\sigma}P_1(\sigma,t) + D(t)\partial_{\sigma}^2P_1(\sigma,t) \\ & \quad \quad - \frac{\theta(|\sigma| - \sigma_{1,c})}{\tau}P_1(\sigma,t) + \delta(\sigma -\sigma_R)\phi\Gamma(t).
\end{split}
\end{equation}
This modification introduces a preferred post-yield stress state in the active population, and therefore acts as a symmetry-breaking field in stress space. In particular, the transformation $(\sigma, \dot{\gamma}) \to (-\sigma, -\dot{\gamma})$ is no longer a symmetry of the microscopic dynamics for fixed $\sigma_R$. However, the full model retains an extended symmetry under simultaneous inversion $(\sigma, \dot{\gamma}, \sigma_R) \to (-\sigma, -\dot{\gamma}, -\sigma_R)$, so that physical observables for negative strain rates can be obtained by symmetry from the positive branch.}

\begin{figure}[b!]
	\includegraphics[width=\columnwidth]{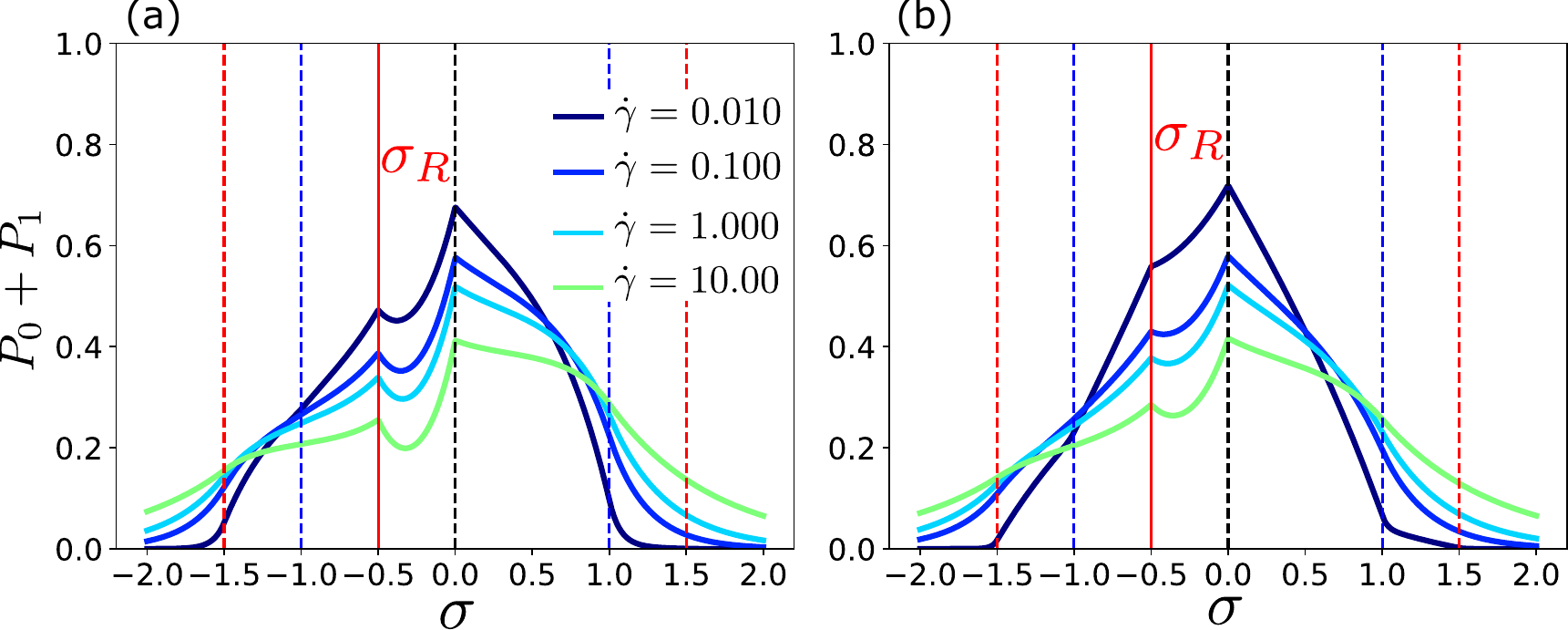}
	\caption{\ai{Probability distributions for $\phi = 0.35$ at (a) $\alpha = 0.4$, and (b) $\alpha = 0.6$. The solid red line indicates the residual stress $\sigma_R = -0.5$, the dashed red lines show the yield stress for active elements $\pm\sigma_{1,c}$, and the dashed blue lines show the yield stress for passive elements $\pm\sigma_{0,c}$. The dashed black line shows $\sigma = 0$.}}
	\label{fig:4}
\end{figure}

\ai{The presence of a residual stress renormalises the yield stress to fluid transition. Solving for the steady state at zero strain rate, $\partial_t P_k = 0$ at zero strain rate, we find a lower threshold value $\alpha_c$ (compared to equation \eqref{eq:alpha_crit_main}) for the yield stress to fluid transition, 
\begin{equation} \label{eq:alpha_crit_res}
\alpha_c = \left[(1 - \phi){\sigma_{0,c}}^2 + \phi\left({\sigma_{1,c}}^2 - \sigma_R^2\right)\right]\big/2.
\end{equation}
The residual stress therefore acts to lower the effective yield threshold, promoting fluidisation of the material. Physically, this reflects the fact that yielding events no longer fully relax stress, leaving a persistent bias that enhances subsequent plastic activity.}

\begin{figure*}[t!]
	\centering
	\includegraphics[width=1.0\textwidth]{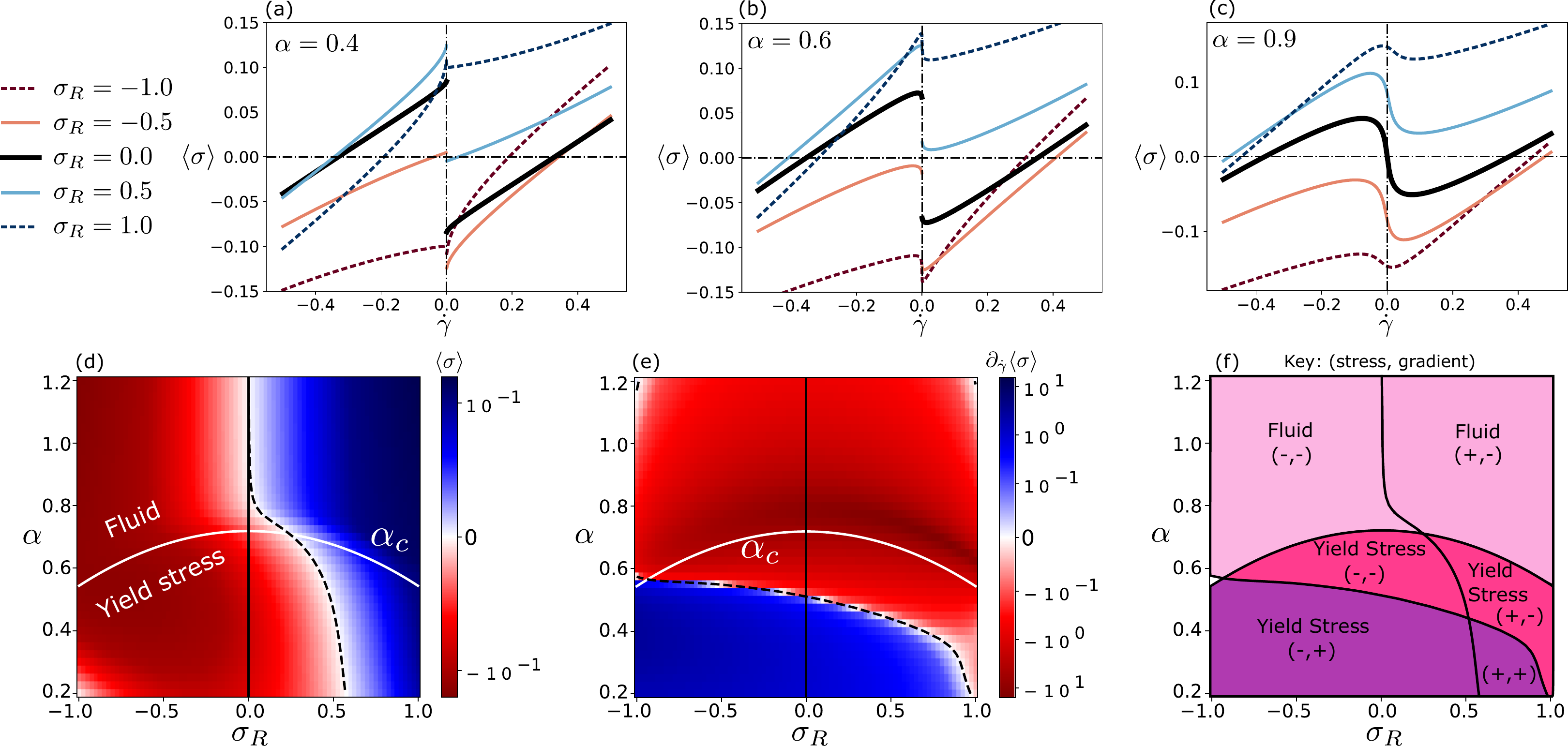}
	\caption{\ai{Flow curves $\langle \sigma \rangle(\dot{\gamma})$ for $\phi = 0.35$ with residual stress for (a) $\alpha = 0.4$, (b) $\alpha = 0.6$, (c) $\alpha = 0.9$. For $\sigma_R = 0$ each of these curves reduces to the corresponding turnover phase from FIG. \ref{fig:3} (d) is shown in sold black. (d) Macroscopic stress $\langle \sigma \rangle$ at low strain rate $\dot{\gamma} = 5 \times 10^{-4}$, with the solid white line indicating the yield-stress/fluid threshold $\alpha_c(\sigma_R)$. The vertical black line shows $\sigma_R = 0$, and the black dashed line shows $\langle \sigma \rangle = 0$. (e) Gradient of the macroscopic stress with respect to strain rate at low strain rate $\dot{\gamma} = 5 \times 10^{-4}$, with the solid white line indicating the yield-stress/fluid threshold $\alpha_c(\sigma_R)$. The vertical black line shows $\sigma_R = 0$, and the black dashed line shows $\partial_{\dot{\gamma}}\langle \sigma \rangle = 0$. (f) Phase diagram indicating phases classified according to both the sign of the macroscopic stress in (a) and its gradient in (b).}}
	\label{fig:6}
\end{figure*}

\ai{A further consequence of the broken symmetry in stress space is the emergence of a finite macroscopic stress at zero strain rate in the fluid phase. For $\alpha > \alpha_c$, the macroscopic stress satisfies
\begin{equation} \label{eq:sigma_0}
\langle\sigma\rangle_0 = \frac{\phi\Gamma\sigma_R\left[s_1\left( 6D_0\tau - \sigma_R^2\right) + s_3\sigma_{1,c}^2 \,\right]}{6s_1D_0},
\end{equation}
where $s_n = \sigma_{1,c} + n\sqrt{D_0\tau}$. Furthermore, in the yield stress phase, the approach to zero strain rate depends on the direction of the limit: $|\langle\sigma\rangle(0^+)| \neq |\langle\sigma\rangle(0^-)|$, while the full symmetry of the model ensures consistency between the two branches via simultaneous inversion of $\sigma, \dot{\gamma}$, and $\sigma_R$.}

\ai{We will focus on the effect of residual stress on the turnover phase \sh{corresponding to the biologically plausible region for convergence-extension}, and fix $\phi = 0.35$. The residual stress does not modify the asymptotic large-strain-rate behaviour, which remains passive for $\phi < 0.5$. At low $\sigma_R$, all three phases FT, YT-, and YT+, remain qualitatively similar to the $\sigma_R = 0$ case (FIG. \ref{fig:6} (a) - (c)). We still observe active behaviour at low strain rate and passive behaviour and large strain rate, however the fluid phase now see a non-zero macroscopic stress $\langle \sigma \rangle(0)$ (equation \eqref{eq:sigma_0}), and the magnitude of the yield stress now depends on whether we approach zero strain rate from the positive or negative side i.e. $|\langle\sigma\rangle(0^+)| \neq |\langle\sigma\rangle(0^-)|$. This is also true for the gradient of the stress at zero strain rate. At higher values of residual stress, the broken symmetry in strain rate is made more apparent.}
\sh{This symmetry breaking is suggestive of a hysteresis mechanism, if the sign of $\sigma_R$ is set by the rheological history of the material, as is plausible through the fabric of the material \cite{guirao2015unified}. For example, increasing the strain rate from negative to positive would yield one branch of the response, and reversing the treatment the other, yielding a cyclic flow curve.}
\sh{We quantify this complexity of flow curves through the stress at $\dot \gamma=0$, $\langle \sigma \rangle(0^+)$ (FIG. \ref{fig:6} (d)), and the initial viscosity $\eta_0 = \lim_{\dot{\gamma} \rightarrow 0^+}\frac{\partial \langle \sigma \rangle}{\partial \dot{\gamma}}$ (FIG. \ref{fig:6} (e)). Finally, we classify the flow by the existence of a yield stress through eq. \ref{eq:alpha_crit_res}. FIG. \ref{fig:6} (f) summarises our findings according to the sign of stress and viscosity (e.g. $(-,+)$), and the presence of a yield stress. }

\section{Discussion} 
We have shown that a two component mean field model produces flow curves that are characteristic of convergence-extension flows found in tissue - active response at low strain rate and passive response at large strain rate. Whether the response at large strain rate is active or passive is determined purely by the active fraction $\phi$. Our model also has a transition from a fluid to a yield stress material, and we report negative yield stresses at large enough values of $\phi$. 
\sh{Allowing for a residual active stress further multiplies the flow curves that can be generated by the model, including states with overall positive or negative shear stress even in the absence of external strain. It should be noted that the active fraction $\phi$ is itself a dynamical quantity that depends on the developmental point of the tissue and, at least in microscopic simulations, its history of applied deformations \cite{sknepnek2021generating,claussen2024geometric,claussen2024mean}.}

\sh{The model developed here should be seen as a first step in understanding the rich rheological phenomenology that is allowed by active elements. It is difficult to translate directly to biological experiment due to its deliberate simplicity. Nevertheless, key parameters can be mapped as follows: The microscopic passive shear modulus $G_{0}$ corresponds to the shear modulus of small groups of cells at low $\dot{\gamma}$, of the order of $1kPa$. The active stress response, $\sigma_a = \sigma_R - |G_1| \gamma$ corresponds to the first two terms in a series expansion in applied strain $\gamma$ of the active response of a quartet of cells undergoing a T1. It should be noted that viscoelastic effects leading to the relaxation of junctions and cell deformations are beyond the scope of our model. The passive and the active yield stress and strain can in principle be extracted from simultaneous measurements of both in e.g. Drosophila germ band extension \cite{brauns2024geometric}. Together, the microscopic stress-strain curves sets the general stress scale $\sigma_Y \approx 1kPa$. The time scale $\tau$ of a T1 transition is approximately 15 minutes \cite{rauzi2008nature,rozbicki2015myosin, brauns2024geometric}, which sets the rate $\tau^{-1}$ in which the strain rate $\dot{\gamma}$ is measured. The dimensionless stress diffusion rate $\alpha$ determines the existence of a yield stress, and both behaviours have been observed in tissues.}

Non-monotonic flow curves such as the ones shown here are responsible for rheological instabilities such as the creation of shear bands \cite{schall2010shear}. As our system is active, it is noteworthy that $\dot \gamma$ is not their main driver, and it can instead act more like a mechanical cue, setting the direction  of potentially complex flow patterns \cite{ioratim2023mechano}. \sh{Together, these observations suggest that active rheology could be a novel driver of developmental patterning, such as in primitive streak formation \cite{rozbicki2015myosin} or in the spatially controlled jamming / unjamming transition in somitogenesis \cite{mongera2018fluid}. A full rheological description of cellular flows needs to take into account cell elongation and orientation, as well as divisions and extrusions in addition to T1s and their orientation \cite{etournay2015interplay,guirao2015unified}.}
A natural next step for elastoplastic models of convergence-extension will therefore be to incorporate active T1s into fully tensorial and spatially extended simulations \cite{liu2018creep,nicolas2018deformation}, as are being currently developed for other active processes such as cell division and extrusion \cite{tahaei2025cell,tahaei2026celldivisionssuppressdynamical}.

\sh{The observations above suggest} that convergence-extension shares features with metamaterials, which exploit the local mechanics to generate auxetic, odd and other responses that can serve as a starting point to engineered functionality \cite{zaiser2023disordered,dudek2025shape}.
\sh{However, the biological version fully rearranges and then rebuilds its mechanical structure, a feat yet to be achieved by engineered systems. If such an active, reloadable element could be developed, e.g. through designing reconfigurable junctions between elements, the model presented here can serve as a road map for a `build your own' rheological response curve. Regions with a negative stress response are the equivalent of a distributed material motor which generates macroscopic stress in response to, and opposite, a deformation. As this behaviour persists into the liquid phase, it could potentially lead to the development of autonomously reconfiguring, liquid engines, mirroring the tissue's mechanobiological function during convergence-extension.}

\begin{acknowledgments}
AIU acknowledges the support of NC3Rs (project reference NC/X002268/1), and EPSRC grant EP/T031077/1. AIU acknowledges the support of the Max Planck Institute. TBL was supported by the UKRI-funded Synthetic Biology Research Centre BrisSynBio (BB/L01386X/1). TBL acknowledges the support of EPSRC grants EP/R014604/1 and EP/T031077/1. 
\end{acknowledgments}

\vspace{-0.5cm}

\bibliography{weird_rheology} 

\clearpage
\onecolumngrid  

\setcounter{equation}{0}  
\setcounter{figure}{0}    
\setcounter{table}{0}     

\renewcommand{\theequation}{S\arabic{equation}} 
\renewcommand{\thefigure}{S\arabic{figure}}
\renewcommand{\thetable}{S\arabic{table}}

\begin{center}
\textbf{\large Exotic rheology of materials with active rearrangements: Supplemental Material}
\end{center}

\section{Stationary solution at zero strain rate}

In order to solve the equations $\partial_tP_k = 0$ ($k = 0, 1$), we split the domain $\sigma$ into four different regions: $-\infty < \sigma \leq -\sigma_{k,c}, -\sigma_{k,c} < \sigma < 0, 0 \leq \sigma < \sigma_{k,c}, \text{ and }\sigma_{k,c} \leq \sigma < \infty$. We require $P_k(0^+) = P_k(0^-)$, the continuity of $P_k(\sigma)$ and it's derivatives at $\sigma = \pm\sigma_{k,c}$, and we require $\lim_{\sigma \to \pm\infty}P_{k}(\sigma) = 0$. When the strain rate is zero, the solutions are
\begin{subequations}
\begin{equation}
P_0(\sigma < -\sigma_{0,c}) = \frac{(1 - \phi)\Gamma}{2}\sqrt{\frac{\tau}{D}}e^{\frac{\sigma + \sigma_{0,c}}{\sqrt{D\tau}}},
\end{equation}
\begin{equation}
P_0(-\sigma_{0,c} < \sigma < \sigma_{0,c}) = \frac{(1 - \phi)\Gamma}{2D}\left[(\theta(-\sigma) - \theta(\sigma))\sigma + \sqrt{D\tau} + \sigma_{0,c}\right],
\end{equation}
\begin{equation}
P_0(\sigma > \sigma_{0,c}) = \frac{(1 - \phi)\Gamma}{2}\sqrt{\frac{\tau}{D}}e^{\frac{\sigma_{0,c} \ai{-\sigma}}{\sqrt{D\tau}}},
\end{equation}
\end{subequations}
for the passive population, and
\begin{subequations}
\begin{equation}
P_1(\sigma < -\sigma_{1,c}) = \frac{\phi\Gamma}{2}\sqrt{\frac{\tau}{D}}e^{\frac{\sigma + \sigma_{1,c}}{\sqrt{D\tau}}},
\end{equation}
\begin{equation}
P_1(-\sigma_{1,c} < \sigma < \sigma_{1,c}) = \frac{\phi\Gamma}{2D}\left[(\theta(-\sigma) - \theta(\sigma))\sigma + \sqrt{D\tau} + \sigma_{1,c}\right],
\end{equation}
\begin{equation}
P_1(\sigma > \sigma_{1,c}) = \frac{\phi\Gamma}{2}\sqrt{\frac{\tau}{D}}e^{\frac{\sigma_{1,c} \ai{-\sigma}}{\sqrt{D\tau}}},
\end{equation}
\end{subequations}
for the active population. These are symmetric, therefore $\langle \sigma \rangle = 0$. The function that determines $D$ is
\begin{equation} \label{eq:alpha_func_zero}
f_{\alpha}\left(\sqrt{D\tau},0,0\right) = D\tau + \sqrt{D\tau}\sigma_{c}^{\text{eff}} + \alpha_c,
\end{equation}
where
\begin{subequations}
\begin{equation} \label{eq:sigma_eff}
\sigma_c^{\text{eff}} = (1 - \phi)\sigma_{0,c} + \phi\sigma_{1,c},
\end{equation}
\begin{equation} \label{eq:alpha_crit}
\alpha_c = \frac{(1 - \phi){\sigma_{0,c}}^2}{2} + \frac{\phi{\sigma_{1,c}}^2}{2}.
\end{equation}
\end{subequations}
Solving the equation $f_{\alpha} = \alpha$, we find that $D(\dot{\gamma} = 0) \equiv D_0$ is given by
\begin{equation} \label{eq:D_zero}
\sqrt{D_0\tau} = \frac{\sigma^{\text{eff}}_c}{2}\bigg[-1 + \sqrt{\frac{4(\alpha - \alpha_c)}{{\sigma^{\text{eff}}_c}^2} + 1}\bigg].
\end{equation}
As in the original HL model, this equation only has a solution for $\alpha > \alpha_c$.

\section{Stationary solution at finite shear rate}

At finite shear rate, we will focus on the case $\dot{\gamma} > 0$ (we have $G_0 > 0$ and $G_1 < 0$). We find the probability distributions to be
\begin{subequations}
\begin{equation}
P_{0}(-\infty < \sigma < -\sigma_{0,c}) = 
\frac{2(1 - \phi)\Gamma\tau}{G_0\dot{\gamma}\tau}\left[\frac{e^{\tfrac{G_0\dot{\gamma}\tau}{2D\tau}(1 + \sqrt{\xi_0})\sigma}}{(1 + \sqrt{\xi_0})e^{\tfrac{G_0\dot{\gamma}\tau}{2D\tau}(1 - \sqrt{\xi_0})\sigma_{0,c}} - (1 - \sqrt{\xi_0})e^{\tfrac{-G_0\dot{\gamma}\tau}{2D\tau}(1 + \sqrt{\xi_0})\sigma_{0,c}}}\right] 
\end{equation}
\begin{equation}
P_{0}(-\sigma_{0,c} < \sigma < 0) = 
\frac{\tfrac{(1 - \phi)\Gamma\tau}{G_0\dot{\gamma}\tau}(1 + \sqrt{\xi_0})e^{
\tfrac{-G_0\dot{\gamma}\tau\sigma_{0,c}}{2D\tau}}\left(e^{\tfrac{G_0\dot{\gamma}\tau(\sigma + \sigma_{0,c})}{D\tau}} + \tfrac{1 - \sqrt{\xi_0}}{1 + \sqrt{\xi_0}}\right)}{(1 + \sqrt{\xi_0})e^{\tfrac{G_0\dot{\gamma}\tau\sigma_{0,c}}{2D\tau}} - (1 - \sqrt{\xi_0})e^{\tfrac{-G_0\dot{\gamma}\tau\sigma_{c}}{2D\tau}}} 
\end{equation}
\begin{equation}
P_{0}(0 < \sigma < \sigma_{0,c}) = 
\frac{\tfrac{(1 - \phi)\Gamma \tau}{G_0\dot{\gamma}\tau}(1 - \sqrt{\xi_0})e^{
\tfrac{G_0\dot{\gamma}\tau\sigma_{0,c}}{2D\tau}}\left(e^{\tfrac{G_0\dot{\gamma}\tau(\sigma - \sigma_{0,c})}{D\tau}} + \tfrac{1 + \sqrt{\xi_0}}{1 -\sqrt{\xi_0}}\right)}{(1 + \sqrt{\xi_0})e^{\tfrac{G_0\dot{\gamma}\tau\sigma_{0,c}}{2D\tau}} - (1 - \sqrt{\xi_0})e^{\tfrac{-G_0\dot{\gamma}\tau\sigma_{0,c}}{2D\tau}}}
\end{equation}
\begin{equation}
P_{0}(\sigma > \sigma_{0,c}) = 
\frac{2(1 - \phi)\Gamma\tau}{G_0\dot{\gamma}\tau}\left[\frac{e^{\tfrac{G_0\dot{\gamma}\tau}{2D\tau}(1 - \sqrt{\xi_0})\sigma}}{(1 + \sqrt{\xi_0})e^{\tfrac{G_0\dot{\gamma}\tau}{2D\tau}(1 - \sqrt{\xi_0})\sigma_{0,c}} - (1 - \sqrt{\xi_0})e^{\tfrac{-G_0\dot{\gamma}\tau}{2D\tau}(1 + \sqrt{\xi_0})\sigma_{0,c}}}\right] 
\end{equation}
\end{subequations}
where 
\begin{equation*}
\xi_0 = 1 + \frac{4D\tau}{(G_0 \dot{\gamma} \tau)^2},
\end{equation*}
for the passive population, and 
\begin{subequations} \label{eq:pact0}
\begin{equation}
P_{1}(-\infty < \sigma < -\sigma_{1,c}) = 
\frac{2\phi\Gamma \tau}{G_1\dot{\gamma}\tau}\left[\frac{e^{\tfrac{G_1\dot{\gamma}\tau}{2D\tau}(1 - \sqrt{\xi_1})\sigma}}{(1 - \sqrt{\xi_1})e^{\tfrac{G_1\dot{\gamma}\tau}{2D\tau}(1 + \sqrt{\xi_1})\sigma_{1,c}} - (1 + \sqrt{\xi_1})e^{\tfrac{-G_1\dot{\gamma}\tau}{2D\tau}(1 - \sqrt{\xi_1})\sigma_{1,c}}}\right] 
\end{equation}
\begin{equation}
P_{1}(-\sigma_{1,c} < \sigma < 0) = 
\frac{\tfrac{\phi\Gamma \tau}{G_1\dot{\gamma}\tau}(1 - \sqrt{\xi_1})e^{
\tfrac{-G_1\dot{\gamma}\tau\sigma_{1,c}}{2D\tau}}\left(e^{\tfrac{G_1\dot{\gamma}\tau(\sigma + \sigma_{1,c})}{D\tau}} + \tfrac{1 + \sqrt{\xi_1}}{1 - \sqrt{\xi_1}}\right)}{(1 - \sqrt{\xi_1})e^{\tfrac{G_1\dot{\gamma}\tau\sigma_{1,c}}{2D\tau}} - (1 + \sqrt{\xi_1})e^{\tfrac{-G_1\dot{\gamma}\tau\sigma_{c}}{2D\tau}}} 
\end{equation}
\begin{equation}
P_{1}(0 < \sigma < \sigma_{1,c}) = 
\frac{\tfrac{\phi\Gamma \tau}{G_1\dot{\gamma}\tau}(1 + \sqrt{\xi_1})e^{
\tfrac{G_1\dot{\gamma}\tau\sigma_{1,c}}{2D\tau}}\left(e^{\tfrac{G_1\dot{\gamma}\tau(\sigma - \sigma_{1,c})}{D\tau}} + \tfrac{1 - \sqrt{\xi_1}}{1 + \sqrt{\xi_1}}\right)}{(1 - \sqrt{\xi_1})e^{\tfrac{G_1\dot{\gamma}\tau\sigma_{1,c}}{2D\tau}} - (1 + \sqrt{\xi_1})e^{\tfrac{-G_1\dot{\gamma}\tau\sigma_{1,c}}{2D\tau}}}
\end{equation}
\begin{equation}
P_{1}(\sigma > \sigma_{1,c}) = 
\frac{2\phi\Gamma \tau}{G_1\dot{\gamma}\tau}\left[\frac{e^{\tfrac{G_1\dot{\gamma}\tau}{2D\tau}(1 + \sqrt{\xi_1})\sigma}}{(1 - \sqrt{\xi_1})e^{\tfrac{G_1\dot{\gamma}\tau}{2D\tau}(1 + \sqrt{\xi_1})\sigma_{1,c}} - (1 + \sqrt{\xi_1})e^{\tfrac{-G_1\dot{\gamma}\tau}{2D\tau}(1 - \sqrt{\xi_1})\sigma_{1,c}}}\right] 
\end{equation}
\end{subequations}
where 
\begin{equation*}
\xi_1 = 1 + \frac{4D\tau}{(G_1 \dot{\gamma} \tau)^2}.
\end{equation*}

\begin{figure*}[t!]
	\centering
	\includegraphics[width=1.0\textwidth]{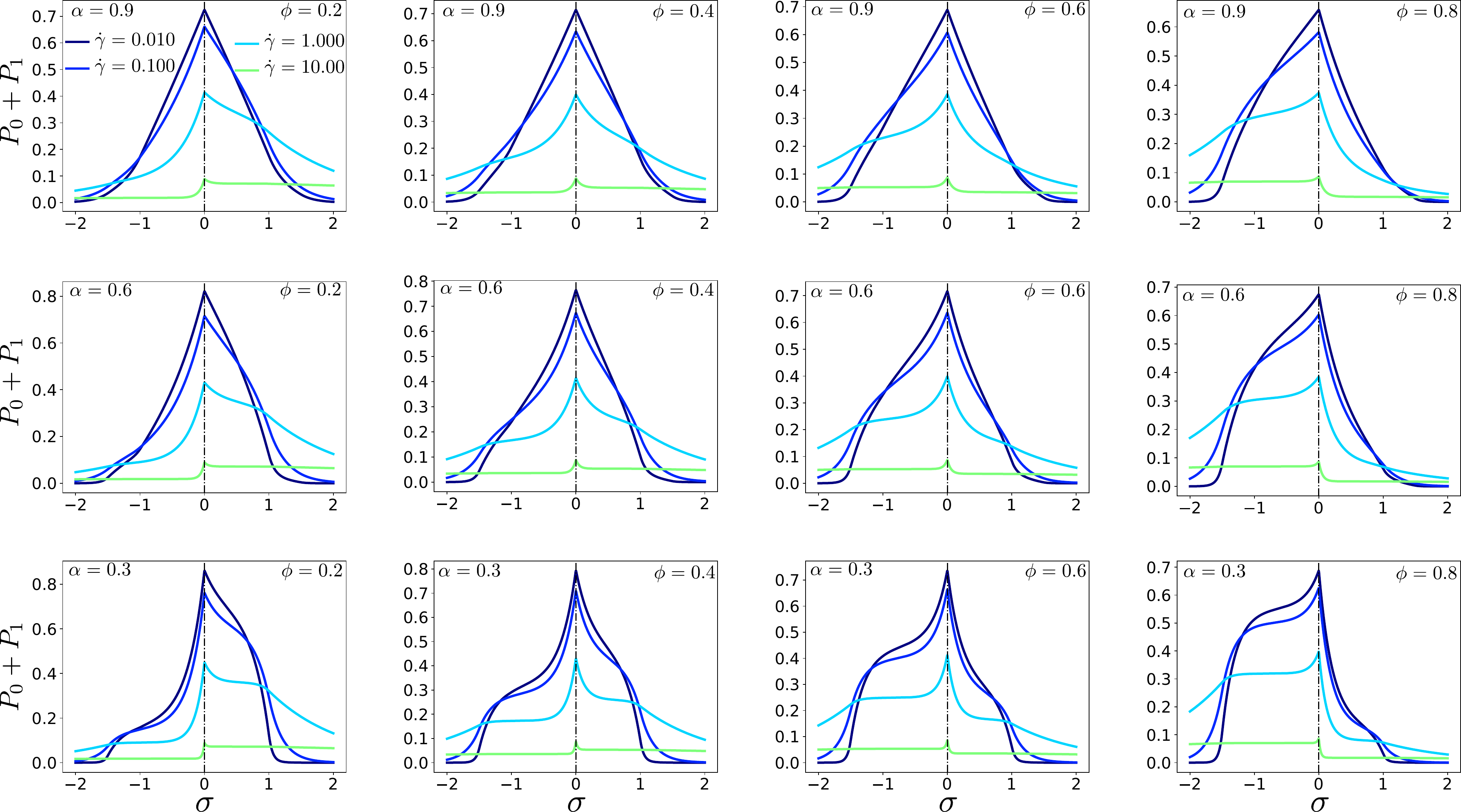}
	\caption{Probability distribution $P_0(\sigma) + P_1(\sigma)$ for $\phi = 0.2, 0.4, 0.6, 0.8$ (left to right) and $\alpha = 0.9, 0.6, 0.3$ (top to bottom). Each curve is a different value of strain rate: $\dot{\gamma} = 0.010, 0.100, 1.000, 10.00$.}
	\label{SI_fig:1}
\end{figure*}

The function that determines $D$ is 
\begin{equation} \label{eq:alpha_func}
\begin{split}
f_{\alpha}\left(\sqrt{D\tau},G_0\dot{\gamma},G_1\dot{\gamma}\right) = D\tau & + \frac{(1 - \phi) \sigma_{0,c}D\tau}{G_0\dot{\gamma}\tau}\frac{1 + \left(\sqrt{\xi_0} + \tfrac{2 D\tau}{\sigma_{0,c}G_0\dot{\gamma}\tau}\right)\tanh{\left(\tfrac{G_0\dot{\gamma}\tau\sigma_{0,c}}{2D\tau}\right)}}{\tanh{\left(\tfrac{G_0\dot{\gamma}\tau\sigma_{0,c}}{2D\tau}\right)} +  \sqrt{\xi_0}} \\ & + \frac{\phi\sigma_{1,c}D\tau}{|G_1|\dot{\gamma}\tau}\frac{1 + \left(\sqrt{\xi_1} + \tfrac{2 D\tau}{\sigma_{1,c}|G_1|\dot{\gamma}\tau}\right)\tanh{\left(\tfrac{|G_1|\dot{\gamma}\tau\sigma_{1,c}}{2D\tau}\right)}}{\tanh{\left(\tfrac{|G_1|\dot{\gamma}\tau\sigma_{1,c}}{2 D\tau}\right)} +  \sqrt{\xi_1}},
\end{split}
\end{equation}
where
\begin{equation*}
\xi_i = 1 + \frac{4D\tau}{\tau^2 G_i^2\dot{\gamma}^2}.
\end{equation*}
The mean stress is 
\begin{equation} \label{eq:mean_stress}
\begin{split}
\langle \sigma \rangle = & \frac{(1 - \phi)D\tau}{f_{\alpha} G_0\dot{\gamma}\tau}\left[f_{\alpha,0}\left(\frac{G_0^2\dot{\gamma}^2\tau^2}{D\tau} - 1 \right) + \frac{2\sigma_{0,c}D\tau}{G_{0}\dot{\gamma}\tau\left(\tanh\left(\frac{G_0\dot{\gamma}\tau\sigma_{0,c}}{2D\tau}\right) + \sqrt{\xi_0}\right)} + \frac{\sigma_{0,c}^2}{2} + D\tau\right] \\ & -\frac{\phi D\tau}{f_{\alpha}|G_1|\dot{\gamma}\tau}\left[f_{\alpha,1}\left(\frac{|G_1|^2\dot{\gamma}^2\tau^2}{D\tau} - 1 \right) + \frac{2\sigma_{1,c}D\tau}{|G_{1}|\dot{\gamma}\tau\left(\tanh\left(\frac{|G_1|\dot{\gamma}\tau\sigma_{1,c}}{2D\tau}\right) + \sqrt{\xi_1}\right)} + \frac{\sigma_{1,c}^2}{2} + D\tau\right]
\end{split}
\end{equation}
where
\begin{subequations}
\begin{equation*}
f_{\alpha,0} = D\tau + \frac{\sigma_{0,c}D\tau}{G_0\dot{\gamma}\tau}\frac{1 + \left(\sqrt{\xi_0} + \tfrac{2 D\tau}{\sigma_{0,c}G_0\dot{\gamma}\tau}\right)\tanh{\left(\tfrac{G_0\dot{\gamma}\tau\sigma_{0,c}}{2D\tau}\right)}}{\tanh{\left(\tfrac{G_0\dot{\gamma}\tau\sigma_{0,c}}{2D\tau}\right)} +  \sqrt{\xi_0}},
\end{equation*}
\begin{equation*}
f_{\alpha,1} = D\tau + \frac{\sigma_{1,c}D\tau}{|G_1|\dot{\gamma}\tau}\frac{1 + \left(\sqrt{\xi_1} + \tfrac{2 D\tau}{\sigma_{1,c}|G_1|\dot{\gamma}\tau}\right)\tanh{\left(\tfrac{|G_1|\dot{\gamma}\tau\sigma_{1,c}}{2D\tau}\right)}}{\tanh{\left(\tfrac{|G_1|\dot{\gamma}\tau\sigma_{1,c}}{2 D\tau}\right)} +  \sqrt{\xi_1}}.
\end{equation*}
\end{subequations}

\section{Macroscopic stress at large shear rate}

We can work out a condition on $G_{1}$ and $G_1$ for $\langle \sigma \rangle$ to be increasing with $\dot{\gamma}$ as $\dot{\gamma} \to \infty$. To take this limit, we first have to figure out from equation \eqref{eq:alpha_func} how $D$ varies with $\dot{\gamma}$. Taking $\dot{\gamma} \to \infty$, which means that $y_i \to \infty$ (Assuming that $D(\dot{\gamma})$ stays finite as $\dot{\gamma} \to \infty$), we find
\begin{equation}
\lim_{\dot{\gamma} \to \infty}f_{\alpha} = D\tau,
\end{equation}
which yields 
\begin{equation}  \label{eq:D_large_gammadot}
D(\dot{\gamma} \to \infty)\tau = \alpha.
\end{equation}
As a consequence of equation \eqref{eq:D_large_gammadot}, 
\begin{equation*}
\Gamma(\dot{\gamma} \to \infty)\tau = D(\dot{\gamma} \to \infty)/\alpha = 1. 
\end{equation*}
Therefore, at large shear rate, we have
\begin{equation}
\lim_{\dot{\gamma} \to \infty}\langle \sigma \rangle =  \left[(1 - \phi)G_0 - \phi\left|G_1\right|\right]\dot{\gamma}\tau.
\end{equation}
Therefore we need 
\begin{equation} \label{eq:passive_behaviour}
\left|G_1\right| < \frac{(1 - \phi)G_0}{\phi}
\end{equation}
for the stress to be positive at large positive shear rate. If $|G_1| = G_0$, then we can only guarantee passive behaviour at large shear rate for $\phi < 0.5$, i.e when the system is more passive than active. 

\begin{figure*}[t!]
	\centering
	\includegraphics[width=1.0\textwidth]{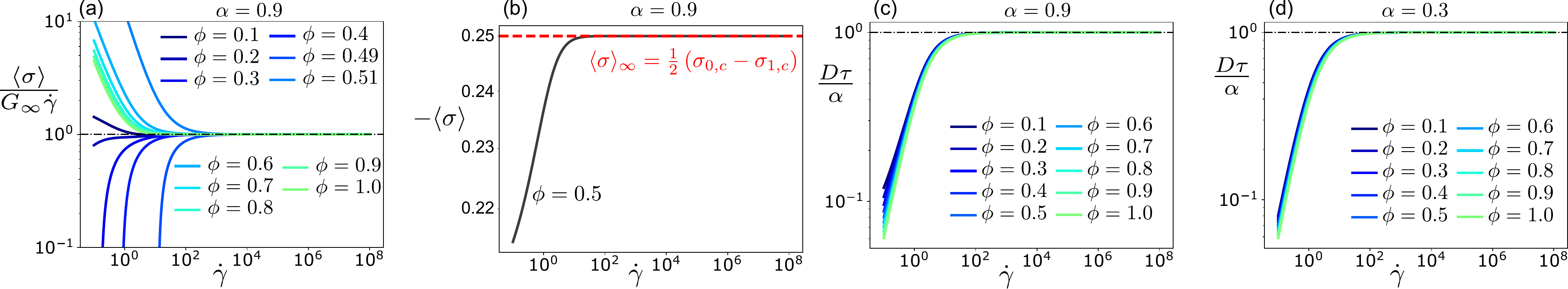}
	\caption{(a) Convergence of the macroscopic stress to $G_{\infty}\dot{\gamma}$. (b) Macroscopic stress as a function of strain rate on a log scale. (c) Convergence to $D\tau$ to $\alpha$ for $\alpha = 0.9$. (d) Convergence to $D\tau$ to $\alpha$ for $\alpha = 0.9$.}
	\label{SI_fig:2}
\end{figure*}

\section{Macroscopic stress at low shear rate}

At low shear rate, we must treat the cases $\alpha < \alpha_c$ and $\alpha > \alpha_c$ separately, where $\alpha_c$ is given by equation \eqref{eq:alpha_crit}. We first need to figure out how $D$ depends on $\dot{\gamma}$ by expanding equation \eqref{eq:alpha_func} for small $\dot{\gamma}$. Following previous work \cite{agoritsas2015relevance}, we expand $D$ in the following way
\begin{equation} \label{eq:D_expansion}
D\tau = 
\begin{cases}
\tilde{C}(1 + \tilde{X}\tau\dot{\gamma}) & \text{for } \alpha > \alpha_c, \\
C\tau\dot{\gamma}(1 + X(\tau\dot{\gamma})^{\frac{1}{2}}) & \text{for } \alpha < \alpha_c.
\end{cases}
\end{equation}
Substituting equation \eqref{eq:D_expansion} into equation \eqref{eq:alpha_func}, we find (for $\dot{\gamma} > 0$)
\begin{equation} \label{eq:expanded_alpha_fuc}
f_{\alpha} =
\begin{cases}
\displaystyle \tilde{C} + \sqrt{\tilde{C}}\sigma_c^{\text{eff}} + \alpha_c + \left[\tilde{C} + \frac{1}{2}\sqrt{\tilde{C}}\sigma_c^{\text{eff}}\right]\tilde{X}\tau\dot{\gamma} + \mathcal{O}(\dot{\gamma}^2) & \text{for } \alpha > \alpha_c, \\
C\left[f\frac{\sigma_{0,c}}{G_0}\tanh\left(\frac{G_0\sigma_{0,c}}{2C}\right) + (1-f)\frac{\sigma_{1,c}}{|G_1|}\tanh\left(\frac{|G_1|\sigma_{1,c}}{2C}\right)\right] + \frac{1}{2}M(C,X)(\tau\dot{\gamma})^{\frac{1}{2}} + \mathcal{O}(\dot{\gamma}) & \text{for } \alpha < \alpha_c,
\end{cases}
\end{equation}
where
\begin{align*}
& M(C,X) = (1 - \phi)\Biggl\{\sqrt{C}\sigma_{0,c} - X\sigma_{0,c}^2 + \left[\frac{2C\sqrt{C}}{G1} + 2CX\sigma_{0,c}\right]\tanh\left(\frac{G_0\sigma_{0,c}}{2C}\right) + \left[X\sigma_{0,c}^2 - \sqrt{C}\sigma_{0,c}\right]\tanh^2\left(\frac{G_0\sigma_{0,c}}{2C}\right)\Biggr\} \\ & + \phi \Biggl\{\sqrt{C}\sigma_{1,c} - X\sigma_{1,c}^2 + \left[\frac{2C\sqrt{C}}{|G_1|} + 2CX\sigma_{1,c}\right]\tanh\left(\frac{|G_1|\sigma_{1,c}}{2C}\right) + \left[X\sigma_{1,c}^2 - \sqrt{C}\sigma_{1,c}\right]\tanh^2\left(\frac{|G_1|\sigma_{1,c}}{2C}\right)\Biggr\}.
\end{align*}
Using $f_{\alpha} = \alpha$, we find
\begin{subequations}
\begin{equation} 
\tilde{C} = D_0\tau,
\end{equation}
\begin{equation}
\tilde{X} = 0,
\end{equation}
\begin{equation}
C\left[(1 - \phi)\frac{\sigma_{0,c}}{G_0}\tanh\left(\frac{G_0\sigma_{0,c}}{2C}\right) + \phi\frac{\sigma_{1,c}}{|G_1|}\tanh\left(\frac{|G_1|\sigma_{1,c}}{2C}\right)\right] = \alpha,
\end{equation}
\begin{equation}
X = \sqrt{C}P(C)/Q(C),
\end{equation}
\end{subequations}
where 
\begin{subequations}
\begin{equation}
\begin{split}
P(C) = & (1 - \phi)\Biggl\{\frac{\sigma_{0,c}}{2} + \frac{C}{G_0}\tanh\left(\frac{G_0\sigma_{0,c}}{2C}\right) - \frac{\sigma_{0,c}}{2}\tanh^2\left(\frac{G_0\sigma_{0,c}}{2C}\right)\Biggr\} \\ & + \phi\Biggl\{\frac{\sigma_{1,c}}{2} + \frac{C}{|G_1|}\tanh\left(\frac{|G_1|\sigma_{1,c}}{2C}\right) - \frac{\sigma_{1,c}}{2}\tanh^2\left(\frac{|G_1|\sigma_{1,c}}{2C}\right)\Biggr\}
\end{split}
\end{equation}
\begin{equation}
\begin{split}
Q(C) = & (1 - \phi)\Biggl\{\frac{\sigma_{0,c}^2}{2} - \frac{C\sigma_{0,c}}{G_0}\tanh\left(\frac{G_0\sigma_{0,c}}{2C}\right) - \frac{\sigma_{0,c}^2}{2}\tanh^2\left(\frac{G_0\sigma_{0,c}}{2C}\right)\Biggr\} \\ & + \phi\Biggl\{\frac{\sigma_{1,c}^2}{2} - \frac{C\sigma_{1,c}}{|G_1|}\tanh\left(\frac{|G_1|\sigma_{1,c}}{2C}\right) - \frac{\sigma_{1,c}^2}{2}\tanh^2\left(\frac{|G_1|\sigma_{1,c}}{2C}\right)\Biggr\}.
\end{split}
\end{equation}
\end{subequations}
We can now calculate the leading order terms of $\langle\sigma\rangle$ as $\dot{\gamma} \to 0$. We find that below $\alpha_c$ the stress is proportional to $\dot{\gamma}$
\begin{equation}
\begin{split}
\langle\sigma\rangle(\alpha > \alpha_c) \approx & \frac{(1 - \phi)}{f_{\alpha}(\sqrt{D_0\tau},0,0)}\left[D_0\tau + \sigma_{0,c}\sqrt{D_0\tau} + \frac{\sigma_{0,c}^2}{2} + \frac{\sigma_{0,c}^3}{6\sqrt{D_0\tau}} + \frac{\sigma_{0,c}^4}{24D_0\tau}\right]G_0\dot{\gamma}\tau \\ & - \frac{\phi}{f_{\alpha}(\sqrt{D_0\tau},0,0)}\left[D_0\tau + \sigma_{1,c}\sqrt{D_0\tau} + \frac{\sigma_{1,c}^2}{2} + \frac{\sigma_{1,c}^3}{6\sqrt{D_0\tau}} + \frac{\sigma_{1,c}^4}{24D_0\tau}\right]|G_1|\dot{\gamma}\tau,
\end{split}
\end{equation}
where 
\begin{equation}
f_\alpha(\sqrt{D_0\tau},0,0) = D_0\tau + \sqrt{D_0\tau}\sigma_c^{\text{eff}} + \alpha_c.
\end{equation}
Below $\alpha_c$, we find a Herschel-Bulkley law with an exponent of $\frac{1}{2}$
\begin{equation}
\langle\sigma\rangle(\alpha < \alpha_c) \approx \sigma_Y + A(\dot{\gamma}\tau)^{\frac{1}{2}},
\end{equation}
with the yield stress given by
\begin{equation} \label{eq:low_yield_stress}
\sigma_Y = \frac{(1 - \phi)|G_1|\left[\frac{\sigma_{0,c}^2}{2} - \frac{C\sigma_{0,c}}{G_0}\tanh\left(\frac{G_0\sigma_{0,c}}{2C}\right)\right] - \phi G_0\left[\frac{\sigma_{1,c}^2}{2} - \frac{C\sigma_{1,c}}{|G_1|}\tanh\left(\frac{|G_1|\sigma_{1,c}}{2C}\right)\right]}{(1 - \phi)|G_1|\sigma_{0,c}\tanh\left(\frac{G_0\sigma_{0,c}}{2C}\right) + \phi G_0\sigma_{1,c}\tanh\left(\frac{|G_1|\sigma_{1,c}}{2C}\right)},
\end{equation}
and the coefficient of the power law given by
\begin{equation} \label{eq:power_law_coeff}
\begin{split}
A = & \frac{\sqrt{C}}{2J(\phi)}\Biggl\{(1 - \phi)|G_1|\sigma_{0,c}\sech^2\left(\frac{G_0\sigma_{0,c}}{2C}\right)\left[\cosh\left(\frac{G1\sigma_{0,c}}{C}\right) - \frac{C}{G_0\sigma_{0,c}}\sinh\left(\frac{G_0\sigma_{0,c}}{C}\right)
\right] \\ & - \phi G_0\sigma_{1,c}\sech^2\left(\frac{|G_1|\sigma_{1,c}}{2C}\right)\left[\cosh\left(\frac{|G_1|\sigma_{1,c}}{C}\right) - \frac{C}{|G_1|\sigma_{1,c}}\sinh\left(\frac{|G_1|\sigma_{1,c}}{C}\right)\right]\Biggr\} \\ & + \frac{\sqrt{C}R(\phi)}{4J(\phi)S(\phi)}\Biggl\{(1 - \phi)|G_1|\sigma_{0,c}^2\sech^2\left(\frac{G_0\sigma_{0,c}}{2C}\right)\left[\cosh\left(\frac{G_0\sigma_{0,c}}{C}\right) - \frac{4C}{G_0\sigma_{0,c}}\sinh\left(\frac{G_0\sigma_{0,c}}{C}\right) + 3\right] \\ & - \phi G_0\sigma_{1,c}^2\sech^2\left(\frac{|G_1|\sigma_{1,c}}{2C}\right)\left[\cosh\left(\frac{|G_1|\sigma_{1,c}}{C}\right) - \frac{4C}{|G_1|\sigma_{1,c}}\sinh\left(\frac{|G_1|\sigma_{1,c}}{C}\right) + 3\right]\Biggr\} 
\end{split}
\end{equation}
where
\begin{subequations}
\begin{equation}
J(\phi) = (1 - \phi)|G_1|\sigma_{0,c}\tanh\left(\frac{G_0\sigma_{0,c}}{2C}\right) + \phi G_0\sigma_{1,c}\tanh\left(\frac{|G_1|\sigma_{1,c}}{2C}\right),
\end{equation}
\begin{equation}
\begin{split}
S(\phi) = & (1 - \phi)G_0|G_1|\sigma_{0,c}^2\sech^2\left(\frac{G_0\sigma_{0,c}}{2C}\right)\left[1 - \frac{C}{G_0\sigma_{0,c}}\sinh\left(\frac{G_0\sigma_{0,c}}{C}\right)\right] \\ & + \phi G_0|G_1|\sigma_{1,c}^2\sech^2\left(\frac{|G_1|\sigma_{1,c}}{2C}\right)\left[1 - \frac{C}{|G_1|\sigma_{1,c}}\sinh\left(\frac{|G_1|\sigma_{1,c}}{C}\right)\right],
\end{split}
\end{equation}
\begin{equation}
\begin{split}
R(\phi) = & (1 - \phi)G_0|G_1|\sigma_{0,c}\sech^2\left(\frac{G_0\sigma_{0,c}}{2C}\right)\left[1 + \frac{C}{G_0\sigma_{0,c}}\sinh\left(\frac{G_0\sigma_{0,c}}{C}\right)\right] \\ & + \phi G_0|G_1|\sigma_{1,c}\sech^2\left(\frac{|G_1|\sigma_{1,c}}{2C}\right)\left[1 + \frac{C}{|G_1|\sigma_{1,c}}\sinh\left(\frac{|G_1|\sigma_{1,c}}{C}\right)\right].
\end{split}
\end{equation}
\end{subequations}
For $\phi = 0$ these expressions reduce to the results of \cite{agoritsas2015relevance} (section 3.4) for the original Hebraud-Lequeux model.

\section{Additional figures}

\begin{figure*}[h!]
	\centering
	\includegraphics[width=1.0\textwidth]{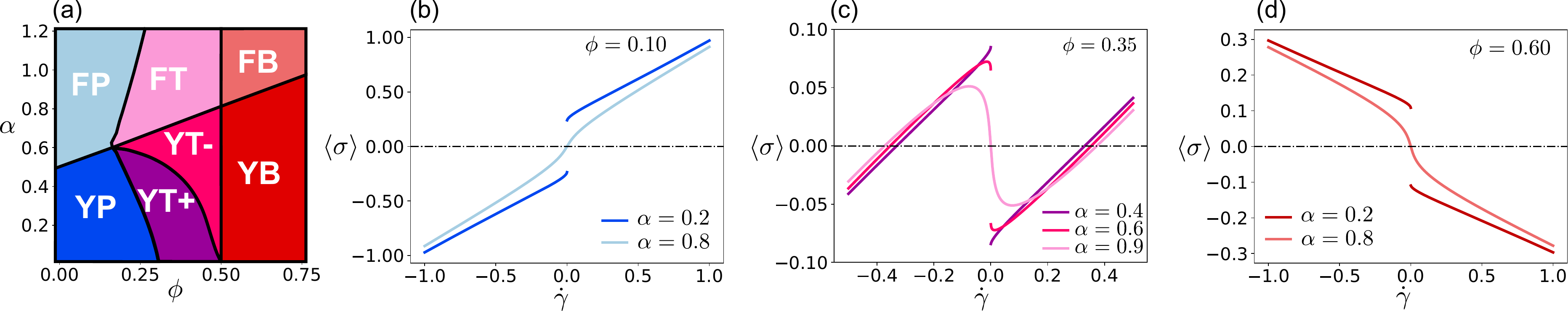}
	\caption{(a) Phase diagram in the $(\phi,\alpha)$ plane constructed from Figs. 3(a) and 3(b) and $G_{\infty}$. There are 6 distinct phases: YB (dark red), FB (light red), YT+ (purple), YT- (pink), FT (light pink), YP (dark blue), and FP (light blue). (b) Typical flow curves for YP (dark blue) and FP (light blue). (c) Typical flow curves for YT+ (purple), YT- (pink), and FT (light pink). (d) Typical flow curves for YB(dark red) and FB (light red).}
	\label{SI_fig:3}
\end{figure*}

\begin{figure*}[h!]
	\centering
	\includegraphics[width=0.6\textwidth]{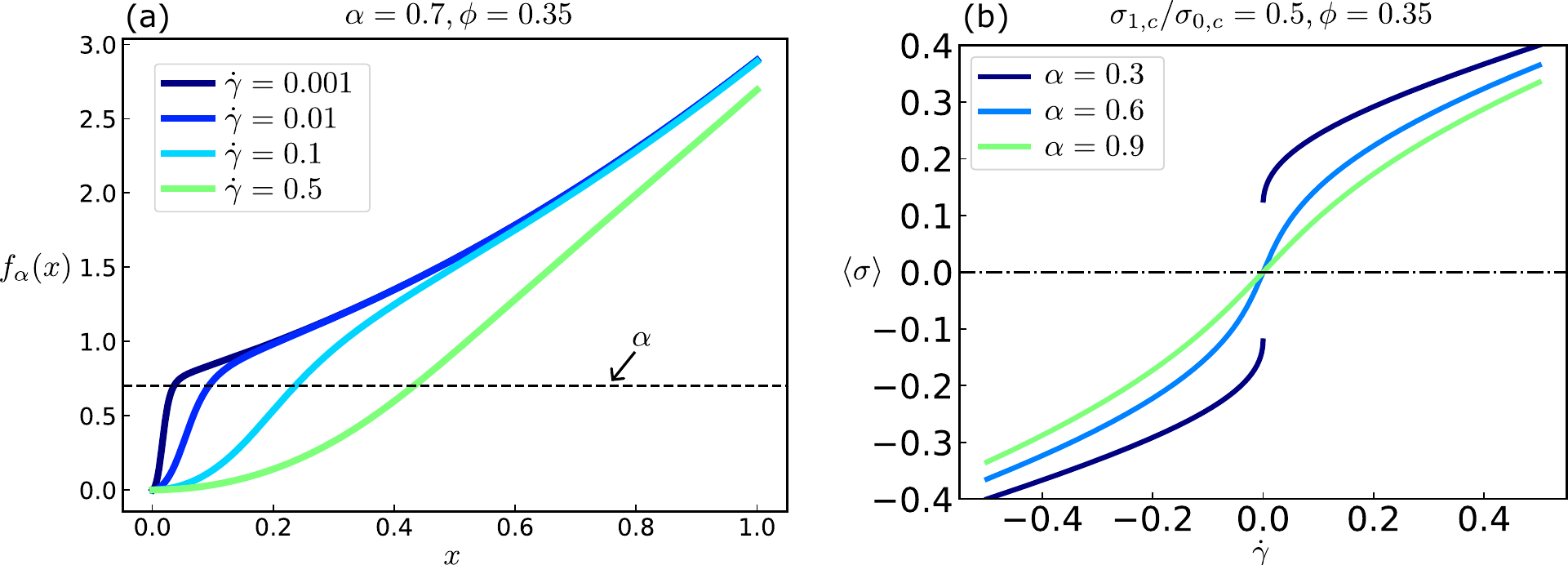}
	\caption{\ai{(a) Graphical representation of the steady state equation $f_{\alpha}(x) = \alpha$ that determines $D$, where $x = \sqrt{D\tau}$. (b) Flow curves for $\sigma_{1,c} < \sigma_{0,c}$, in direct contrast to FIG. 4(d) in the main text where $\sigma_{1,c} < \sigma_{0,c}$ with all other parameters the same.}}
	\label{SI_fig:3.5}
\end{figure*}

\section{Non-zero residual stress}
\begin{figure*}[b!]
	\centering
	\includegraphics[width=1.0\textwidth]{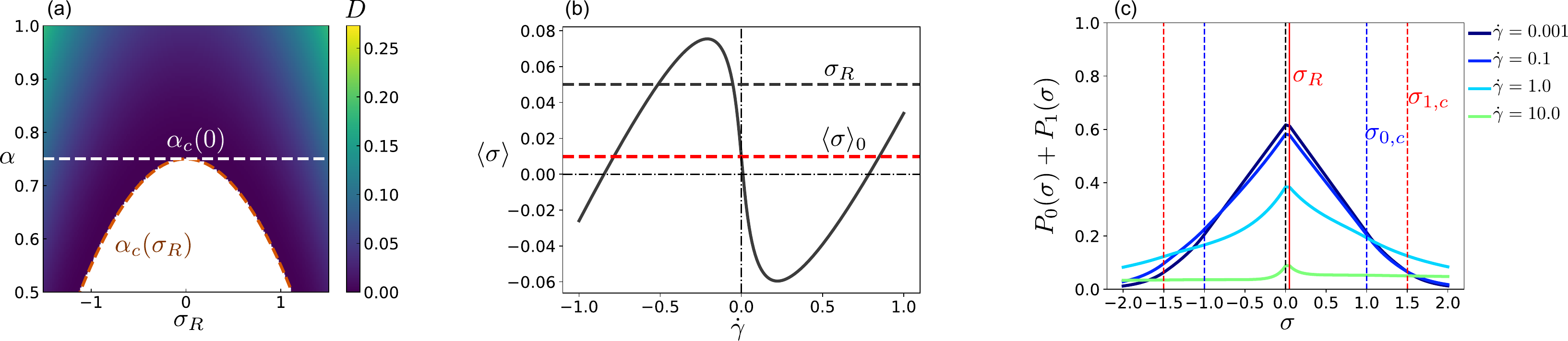}
	\caption{\ai{(a) Numerical solution to \eqref{eq:falpha0} (colour plot) compared with $\alpha_c(\sigma_R)$ (orange dotted line) given by \eqref{eq:alpahc}. (b) Flow curve for $\phi = 0.4$, $\alpha = 1.2$, $\sigma_R = 0.05$, \ai{the red line shows $\langle \sigma \rangle_0$ given by \eqref{eq:sigma0}} and (c) probability distribution for the same parameters.}} 
	\label{RR_fig:1}
\end{figure*}

\ai{
In order to break symmetry and address concerns about the lack of residual stress and PCP symmetry breaking, we update the model in the simplest way by adding a residual stress $\sigma_R$ to the active population:
\begin{equation}
\partial_{t}P_1(\sigma,t) = -G_1\dot{\gamma}\partial_{\sigma}P_1(\sigma,t) + D(t)\partial_{\sigma}^2P_1(\sigma,t) - \frac{\theta(|\sigma| - \sigma_{1,c})}{\tau}P_1(\sigma,t) + \delta(\sigma - \sigma_{R})\phi\Gamma(t),
\end{equation}
where $-\sigma_{1,c} < \sigma_{R} < \sigma_{1,c}$. This modification allows the symmetry under $\sigma \to -\sigma$, $\dot{\gamma} \to -\dot{\gamma}$ to be broken. The probability distribution is piecewise with continuity of $P_1$ and its first derivative at $\sigma = \pm\sigma_{1,c}$. Because yielded elements reset to $\sigma = \sigma_R$, we also have continuity of $P_1$ at $\sigma = \sigma_R$, but \emph{not} its first derivative. Therefore, we can write
\begin{subequations}
\begin{align}
-g\partial_{\sigma} P_1 + D\partial_{\sigma}^2P_1 - \frac{1}{\tau}P_1 = 0, && \sigma < -\sigma_{1,c},
\end{align}
\begin{align} \label{eq:middle_region}
-g\partial_{\sigma} P_1 + D\partial_{\sigma}^2P_1 - \delta\left(\sigma - \sigma_R\right)\phi\Gamma = 0, && -\sigma_{1,c} < \sigma < \sigma_{1,c},
\end{align}
\begin{align}
-g\partial_{\sigma} P_1 + D\partial_{\sigma}^2P_1 - \frac{1}{\tau}P_1 = 0, && \sigma > \sigma_{1,c},
\end{align}
\end{subequations}
where the discontinuity of the first derivative of $P_1$ is determined by integrating \eqref{eq:middle_region} around $\sigma = \sigma_R$.}
\ai{
At zero shear rate, we have
\begin{subequations}
\begin{equation}
P_{1}(-\infty < \sigma < -\sigma_{1,c}) = A^{(1)}e^{\sigma/\sqrt{D\tau}},
\end{equation}
\begin{equation}
P_{1}(-\sigma_{1,c} < \sigma < \sigma_R) = A^{(2)}\sigma + B^{(2)} + \frac{\phi\Gamma}{2D}(\sigma -\sigma_R),
\end{equation}
\begin{equation}
P_{1}(\sigma_R < \sigma < \sigma_{1,c}) = A^{(2)}\sigma + B^{(2)} + \frac{\phi\Gamma}{2D}(\sigma_R -\sigma),
\end{equation}
\begin{equation}
P_{1}(\sigma > \sigma_{1,c}) = A^{(3)}e^{-\sigma/\sqrt{D\tau}},
\end{equation}
\end{subequations}
where the middle two expressions have been obtained by applying continuity and the jump condition either side of $\sigma = \sigma_R$. If we let
\begin{align*}
P(\sigma_R^+) = C^+\sigma + B^+, && P(\sigma_R^-) = C^-\sigma + B^-,
\end{align*}
and enforce continuity plus the the jump condition $D\partial_\sigma P(\sigma_R^+) - D\partial_\sigma P(\sigma_R^+) + \phi\Gamma = 0$, we find
\begin{align*}
C^+ - C^- = -\frac{\phi\Gamma}{D}, && B^- - B^+ = \sigma_R \left(C^+ - C^-\right).
\end{align*}
This forces a choice as we have 2 equations for 4 unknowns, and I will make the symmetric one:
\begin{align*}
B^{+} = B^{(2)} + \frac{\phi\Gamma\sigma_R}{2D}, && B^{-} = B^{(2)} - \frac{\phi\Gamma\sigma_R}{2D}, && C^{+} = A^{(2)} - \frac{\phi\Gamma}{2D}, && C^{-} = A^{(2)} + \frac{\phi\Gamma}{2D}.
\end{align*}
Applying continuity at $\pm\sigma_{1,c}$, we find
\begin{align*}
A^{(1)} = \frac{e^{\sigma_{1,c}/\sqrt{D\tau}}\phi\Gamma\sqrt{D\tau}(\sigma_{1,c} - \sigma_R + \sqrt{D\tau})}{2D(\sigma_{1,c} + \sqrt{D\tau})}, && A^{(3)} = \frac{e^{\sigma_{1,c}/\sqrt{D\tau}}\phi\Gamma\sqrt{D\tau}(\sigma_{1,c} + \sigma_R + \sqrt{D\tau})}{2D(\sigma_{1,c} + \sqrt{D\tau})},
\end{align*}
\begin{align*}
A^{(2)} = -\frac{\phi\Gamma\sigma_R}{2D(\sigma_{1,c} + \sqrt{D\tau})}, && B^{(2)} = \frac{\phi\Gamma(\sigma_{1,c} + \sqrt{D\tau})}{2D}.
\end{align*}
The probability distribution must be normalised, so we require
\begin{align*}
\int_{-\infty}^{\infty} (P_0 + P_1) \, d\sigma = 1,
\end{align*}
and using the definition of the plastic activity $\Gamma$, we find
\begin{equation} \label{eq:normalise}
\Gamma\tau + \int_{-\sigma_{0,c}}^{\sigma_{0,c}} P_0 \, d\sigma + \int_{-\sigma_{1,c}}^{\sigma_{1,c}} P_1 \, d\sigma = 1.
\end{equation}
Combining \eqref{eq:normalise} with self-consistency $D = \alpha\Gamma$, we find 
\begin{equation} \label{eq:falpha0}
D\tau + \sqrt{D\tau}\bigg[(1 - \phi)\sigma_{0,c} + \phi\sigma_{1,c}\bigg] + \frac{(1-\phi)\sigma_{0,c}^2}{2} + \frac{\phi(\sigma_{1,c}^2 - \sigma_R^2)}{2} - \alpha = 0,
\end{equation}
which has solution
\begin{equation} \label{eq:falpha_R}
\sqrt{D_0\tau} = \frac{(1 - \phi)\sigma_{0,c} + \phi\sigma_{1,c}}{2}\left[- 1 + \sqrt{1 + \frac{4(\alpha - \alpha_c)}{\left[(1 - \phi)\sigma_{0,c} + \phi\sigma_{1,c}\right]^2}}\,\right],
\end{equation}
for $\alpha > \alpha_c$ (figure \ref{RR_fig:1})
\begin{equation} \label{eq:alpahc}
\alpha_c = \frac{(1 - \phi)\sigma_{0,c}^2}{2} + \frac{\phi(\sigma_{1,c}^2 - \sigma_R^2)}{2}.
\end{equation}
For $\sigma_R \neq 0$, we now have a non-zero macroscopic stress:
\begin{equation} \label{eq:sigma0}
\langle \sigma \rangle_0 = \frac{\phi\Gamma\sigma_R}{6D_0\left(\sigma_{1,c} + \sqrt{D_0\tau}\right)}\bigg[\sigma_{1,c}^3 - \sigma_{1,c}\sigma_R^2 + 3\sigma_{1,c}^2\sqrt{D_0\tau} - \sigma_R^2\sqrt{D_0\tau} + 6\sigma_{1,c}D_0\tau + 6\sqrt{D_0\tau}^3\bigg].
\end{equation}
}
\ai{
For non-zero shear rate, we use the shorthand $g = G_1\dot{\gamma}$, and we will proceed by assuming $g < 0$ (the solution for $g > 0$ can be obtained by taking $\sqrt{\xi} \to -\sqrt{\xi}$). First we deal with the matching and jump condition due to the Dirac delta at $\sigma = \sigma_R$. Between $-\sigma_{1,c}$ and $\sigma_{1,c}$, we have 
\begin{align*}
P_1(\sigma < \sigma_R) = A^{(2)}_-e^{y\sigma} + B^{(2)}_-, && P_1(\sigma > \sigma_R) = A^{(2)}_+e^{y\sigma} + B^{(2)}_+,
\end{align*}
where $y = g/D$. The jump condition and continuity require
\begin{subequations}
\begin{equation} \label{eq:jump_at_R}
-g\left[P_1^+ - P_1^-\right] + D\left[\partial_{\sigma}P_1^+ - \partial_{\sigma}P_1^-\right] + \phi\Gamma = 0,
\end{equation}
\begin{equation} \label{eq:cont_at_R}
P_1^+ - P_1^- = 0,
\end{equation}
\end{subequations}
which implies $D\left[\partial_{\sigma}P_1^+ - \partial_{\sigma}P_1^-\right] + \phi\Gamma = 0$ when \eqref{eq:cont_at_R} is substituted into \eqref{eq:jump_at_R}. The coefficients therefore must satisfy
\begin{align*}
A^{(2)}_+ - A^{(2)}_- = -\frac{\phi\Gamma e^{-y\sigma_R}}{g}, && B^{(2)}_- - B^{(2)}_+ = -\frac{\phi\Gamma}{g},
\end{align*}
and again we make the symmetric choice
\begin{align*}
A^{(2)}_+ = A^{(2)} - \frac{\phi\Gamma e^{-y\sigma_R}}{2g}, && A^{(2)}_- = A^{(2)} + \frac{\phi\Gamma e^{-y\sigma_R}}{2g}, && B^{(2)}_+ = B^{(2)} + \frac{\phi\Gamma}{2g}, && B^{(2)}_- = B^{(2)} - \frac{\phi\Gamma}{2g}.
\end{align*}
Putting this together, we can write 
\begin{subequations}
\begin{equation}
P_{1}(-\infty < \sigma < -\sigma_{1,c}) = A^{(1)}e^{\frac{y}{2}(1 - \sqrt{\xi})\sigma},
\end{equation}
\begin{equation}
P_{1}(-\sigma_{1,c} < \sigma < \sigma_R) = A^{(2)}e^{y\sigma} + B^{(2)} + \frac{\phi\Gamma}{2g}\bigg[e^{-y\sigma_R}e^{y\sigma} - 1\bigg],
\end{equation}
\begin{equation}
P_{1}(\sigma_{R} < \sigma < \sigma_{1,c}) = A^{(2)}e^{y\sigma} + B^{(2)} - \frac{\phi\Gamma}{2g}\bigg[e^{-y\sigma_R}e^{y\sigma} - 1\bigg],
\end{equation}
\begin{equation}
P_{1}(\sigma > \sigma_{1,c}) = A^{(3)}e^{\frac{y}{2}(1 + \sqrt{\xi})\sigma},
\end{equation}
\end{subequations}
where $y = g/D$. Matching $P_1$ and its first derivatives at $\sigma = \sigma_{1,c}$ give the following four equations for four unknonws $A^{1}, A^{(3)}, A^{(2)}, B^{(2)}$:
\begin{subequations}
\begin{equation} \label{eq:1}
A^{(1)}e^{-\frac{y}{2}(1 - \sqrt{\xi})\sigma_{1,c}} = A^{(2)}e^{-y\sigma_{1,c}} + B^{(2)} + \frac{\phi\Gamma}{2g}\bigg[e^{-y\sigma_R}e^{-y\sigma_c} - 1\bigg],
\end{equation}
\begin{equation} \label{eq:2}
A^{(3)}e^{\frac{y}{2}(1 + \sqrt{\xi})\sigma_{1,c}} = A^{(2)}e^{y\sigma_{1,c}} + B^{(2)} - \frac{\phi\Gamma}{2g}\bigg[e^{-y\sigma_R}e^{y\sigma_c} - 1\bigg],
\end{equation}
\begin{equation} \label{eq:3}
\frac{y}{2}(1 - \sqrt{\xi})A^{(1)}e^{-\frac{y}{2}(1 - \sqrt{\xi})\sigma_{1,c}} = yA^{(2)}e^{-y\sigma_c} + \frac{\phi\Gamma}{2D}e^{-y\sigma_R}e^{-y\sigma_{1,c}},
\end{equation}
\begin{equation} \label{eq:4}
\frac{y}{2}(1 + \sqrt{\xi})A^{(1)}e^{\frac{y}{2}(1 + \sqrt{\xi})\sigma_{1,c}} = yA^{(2)}e^{y\sigma_c} - \frac{\phi\Gamma}{2D}e^{-y\sigma_R}e^{y\sigma_{1,c}}.
\end{equation}
\end{subequations}
Solving equations \eqref{eq:1} - \eqref{eq:4}, we find 
\begin{align*}
A^{(2)} = \frac{\phi\Gamma}{g}\left[\frac{S(\sigma_R)}{4}\left(\frac{\alpha + \beta}{\alpha - \beta} - \frac{\alpha - \beta}{\alpha + \beta}\right) + \frac{e^{-y\sigma_R}}{2}\left(\frac{\alpha + \beta}{\alpha - \beta}\right)\right], && B^{(2)} = \frac{\phi\Gamma}{g}\left[\frac{S(\sigma_R)}{4}\left(\frac{\alpha + \beta}{\alpha - \beta} + \frac{\alpha - \beta}{\alpha + \beta}\right) + \frac{e^{-y\sigma_R}}{2}\left(\frac{\alpha + \beta}{\alpha - \beta}\right)\right],
\end{align*}
\begin{align*}
A^{(1)} = \frac{\phi\Gamma}{g}e^{-\frac{y\sigma_{1,c}\sqrt{\xi}}{2}}\left[\frac{S(\sigma_R)}{2\alpha}\left(\frac{\alpha + \beta}{\alpha - \beta} - \frac{\alpha - \beta}{\alpha + \beta}\right) + \frac{2e^{-y\sigma_R}}{\alpha - \beta}\right], && A^{(3)} = \frac{\phi\Gamma}{g}e^{-\frac{y\sigma_{1,c}\sqrt{\xi}}{2}}\left[\frac{S(\sigma_R)}{2\beta}\left(\frac{\alpha + \beta}{\alpha - \beta} - \frac{\alpha - \beta}{\alpha + \beta}\right) + \frac{2e^{-y\sigma_R}}{\alpha - \beta}\right],
\end{align*}
where 
\begin{align*}
y = \frac{g}{D}, && S(\sigma_R) = 1 - e^{-y\sigma_R}, && \alpha = (1 - \sqrt{\xi})e^{\frac{y}{2}\sigma_{1,c}}, && \beta = (1 + \sqrt{\xi})e^{-\frac{y}{2}\sigma_{1,c}}, && \xi = 1 + 4D\tau/(g\tau)^2.
\end{align*}
Thus the probability distribution for the active population is 
\begin{subequations}
\begin{equation}
P_{1}(-\infty < \sigma < -\sigma_{1,c}) = \frac{\phi\Gamma}{g}e^{-\frac{y}{2}\sigma_{1,c}\sqrt{\xi}}\left[\frac{S(\sigma_R)}{2\alpha}\left(\frac{\alpha + \beta}{\alpha - \beta} - \frac{\alpha - \beta}{\alpha + \beta}\right) + \frac{2e^{-y\sigma_R}}{\alpha - \beta}\right]e^{\frac{y}{2}(1 - \sqrt{\xi})\sigma},
\end{equation}
\begin{equation}
\begin{split}
P_{1}(-\sigma_{1,c} < \sigma < \sigma_R) & = \frac{\phi\Gamma}{g}\left[\frac{S(\sigma_R)}{4}\left(\frac{\alpha + \beta}{\alpha - \beta} - \frac{\alpha - \beta}{\alpha + \beta}\right) + \frac{1}{2}e^{-y\sigma_R}\left(\frac{\alpha + \beta}{\alpha - \beta}\right)\right]e^{y\sigma} + \frac{\phi\Gamma}{2g}\bigg[e^{y\sigma}e^{-y\sigma_R} - 1\bigg] \\ & + \frac{\phi\Gamma}{g}\left[\frac{S(\sigma_R)}{4}\left(\frac{\alpha + \beta}{\alpha - \beta} + \frac{\alpha - \beta}{\alpha + \beta}\right) + \frac{1}{2}e^{-y\sigma_R}\left(\frac{\alpha + \beta}{\alpha - \beta}\right)\right]
\end{split}
\end{equation}
\begin{equation}
\begin{split}
P_{1}(\sigma_{R} < \sigma < \sigma_{1,c}) & = \frac{\phi\Gamma}{g}\left[\frac{S(\sigma_R)}{4}\left(\frac{\alpha + \beta}{\alpha - \beta} - \frac{\alpha - \beta}{\alpha + \beta}\right) + \frac{1}{2}e^{-y\sigma_R}\left(\frac{\alpha + \beta}{\alpha - \beta}\right)\right]e^{y\sigma} - \frac{\phi\Gamma}{2g}\bigg[e^{y\sigma}e^{-y\sigma_R} - 1\bigg] \\ & + \frac{\phi\Gamma}{g}\left[\frac{S(\sigma_R)}{4}\left(\frac{\alpha + \beta}{\alpha - \beta} + \frac{\alpha - \beta}{\alpha + \beta}\right) + \frac{1}{2}e^{-y\sigma_R}\left(\frac{\alpha + \beta}{\alpha - \beta}\right)\right]
\end{split}
\end{equation}
\begin{equation}
P_{1}(\sigma > \sigma_{1,c}) = \frac{\phi\Gamma}{g}e^{-\frac{y}{2}\sigma_{1,c}\sqrt{\xi}}\left[\frac{S(\sigma_R)}{2\beta}\left(\frac{\alpha + \beta}{\alpha - \beta} - \frac{\alpha - \beta}{\alpha + \beta}\right) + \frac{2e^{-y\sigma_R}}{\alpha - \beta}\right]e^{\frac{y}{2}(1 + \sqrt{\xi})\sigma},
\end{equation}
\end{subequations}
The solution for $g > 0$ is obtained by taking $\sqrt{\xi} \to -\sqrt{\xi}$. The total probability distribution $P_0 + P_1$ is plotted in figure \ref{RR_fig:1}.}

\ai{
By enforcing that the total probability distribution be normalised, i.e. $\int_{-\infty}^{\infty}(P_0 + P_1)\,d\sigma = 1$, and requiring self consistency via $D = \alpha\Gamma$, we find that $D$ is the solution to $f_{\alpha} - \alpha = 0$, where
\begin{equation}
\begin{split}
f_{\alpha} & = \frac{\phi}{y}\Bigg\{\sigma_{1,c}\left[\frac{S(\sigma_R)}{2}\left(\frac{\alpha + \beta}{\alpha - \beta} + \frac{\alpha - \beta}{\alpha + \beta}\right) + e^{-y\sigma_R}\left(\frac{\alpha + \beta}{\alpha - \beta}\right)\right] + \frac{\sinh(y\sigma_{1,c})}{y}\left[\frac{S(\sigma_R)}{2}\left(\frac{\alpha + \beta}{\alpha - \beta} - \frac{\alpha - \beta}{\alpha + \beta}\right) + e^{-y\sigma_R}\left(\frac{\alpha + \beta}{\alpha - \beta}\right)\right] \\ & \quad + \frac{1}{y} - \sigma_R - \frac{e^{-y\sigma_R}\cosh(y\sigma_{1,c})}{y}\Bigg\} + D\tau + \frac{(1 - \phi) \sigma_{0,c}D\tau}{G_0\dot{\gamma}\tau}\frac{1 + \left(\sqrt{\xi_0} + \tfrac{2 D\tau}{\sigma_{0,c}G_0\dot{\gamma}\tau}\right)\tanh{\left(\tfrac{G_0\dot{\gamma}\tau\sigma_{0,c}}{2D\tau}\right)}}{\tanh{\left(\tfrac{G_0\dot{\gamma}\tau\sigma_{0,c}}{2D\tau}\right)} +  \sqrt{\xi_0}},
\end{split}
\end{equation}
where $\xi_0 = 1 + 4D/(\tau G_0^2\dot{\gamma}^2)$. In the limit $\dot{\gamma \to 0}$, with $D = C(1 + X\tau\dot{\gamma})$ we find 
\begin{align*}
C + \sqrt{C}\bigg[(1 - \phi)\sigma_{0,c} + \phi\sigma_{1,c}\bigg] + \frac{(1-\phi)\sigma_{0,c}^2}{2} + \frac{\phi(\sigma_{1,c}^2 - \sigma_R^2)}{2} - \alpha = 0,
\end{align*}
which agrees with \eqref{eq:falpha_R}, with $C = D_0\tau$. }
\ai{
The active contribution to the mean stress is given by $\langle \sigma \rangle_1 = \int_{-\infty}^{\infty} \sigma P_1 \, d\sigma$. We can compute this in pieces:
\begin{subequations}
\begin{align*}
\int_{-\infty}^{-\sigma_{1,c}} \sigma P_1(\sigma) \, d\sigma = \frac{-2A^{(1)e^{\frac{-y\sigma_{1,c}}{2}(1 - \sqrt{\xi})}}\left[2 + y\sigma_{1,c}(1 - \sqrt{\xi})\right]}{y^2(1 - \sqrt{\xi})^2},
\end{align*}
\begin{align*}
\int_{\sigma_{1,c}}^{\infty} \sigma P_1(\sigma) \, d\sigma  = \frac{-2A^{(3)e^{\frac{y\sigma_{1,c}}{2}(1 + \sqrt{\xi})}}\left[-2 + y\sigma_{1,c}(1 + \sqrt{\xi})\right]}{y^2(1 + \sqrt{\xi})^2},
\end{align*}
\begin{align*}
\int_{-\sigma_{1,c}}^{\sigma_R} \sigma P_1(\sigma) \, d\sigma + \int_{\sigma_R}^{\sigma_{1,c}} \sigma P_1(\sigma) \, d\sigma & = \frac{e^{-y\sigma_{1,c}}e^{-y\sigma_R}}{2y^2}\Bigg\{2A^{(2)}e^{y\sigma_R}\bigg[1 + y\sigma_{1,c} - e^{2y\sigma_{1,c}} + y\sigma_{1,c}e^{2y\sigma_{1,c}}\bigg] \\ & \quad + \frac{\phi\Gamma}{g}\bigg[1 + y\sigma_{1,c} + e^{2y\sigma_{1,c}} - y\sigma_{1,c}e^{2y\sigma_{1,c}} + e^{y\sigma_{1,c}}e^{y\sigma_R}\left(-2 + y^2\sigma_{1,c}^2 + 2y\sigma_R - y^2\sigma_R^2\right)\bigg]\Bigg\},
\end{align*}
\end{subequations}
where
\begin{align*}
y = \frac{g}{D}, && S(\sigma_R) = 1 - e^{-y\sigma_R}, && \alpha = (1 - \sqrt{\xi})e^{\frac{y}{2}\sigma_{1,c}}, && \beta = (1 + \sqrt{\xi})e^{-\frac{y}{2}\sigma_{1,c}}, && \xi = 1 + 4D\tau/(g\tau)^2.
\end{align*}
After substituting the coefficients and simplifying, we find 
\begin{equation}
\begin{split} 
\langle \sigma \rangle_1 & = \frac{\phi\Gamma}{gy^2}\Bigg\{\left[\frac{S(\sigma_R)}{2}\left(\frac{\alpha + \beta}{\alpha - \beta} -\frac{\alpha - \beta}{\alpha + \beta}\right) + e^{-y\sigma_R}\left(\frac{\alpha + \beta}{\alpha - \beta}\right)\right]\bigg[y\sigma_{1,c}\cosh(y\sigma_{1,c}) - \sinh(y\sigma_{1,c})\bigg] \\ & \quad - \left[\frac{S(\sigma_R)}{\alpha}\left(\frac{\alpha + \beta}{\alpha - \beta} - \frac{\alpha - \beta}{\alpha + \beta}\right) + \frac{4e^{-y\sigma_R}}{\alpha - \beta}\right]\left[\frac{2e^{y\sigma_{1,c}/2}}{\alpha^2} + \frac{y\sigma_{1,c}}{\alpha}\right] \\ & \quad - \left[\frac{S(\sigma_R)}{\beta}\left(\frac{\alpha + \beta}{\alpha - \beta} - \frac{\alpha - \beta}{\alpha + \beta}\right) + \frac{4e^{-y\sigma_R}}{\alpha - \beta}\right]\left[\frac{-2e^{y\sigma_{1,c}/2}}{\beta^2} + \frac{y\sigma_{1,c}}{\beta}\right]  \\ & \quad + e^{-y\sigma_R}\bigg[\cosh(y\sigma_{1,c}) - y\sigma_{1,c}\sinh(y\sigma_{1,c})\bigg] + \frac{1}{2}\bigg[2y\sigma_R - 2 + y^2\left(\sigma_{1,c}^2 - \sigma_R^2\right)\bigg]\Bigg\}.
\end{split}
\end{equation}
We see from FIG. \ref{RR_fig:1} that $ \langle \sigma \rangle(\dot{\gamma} = 0) = \langle \sigma \rangle_0$, given by \eqref{eq:sigma0}.}

\end{document}